%% file: main_v3.tex
\documentclass[CRPHYS,Unicode,manuscript]{cedram}

\usepackage{amsmath}
\usepackage{physics}
\usepackage{amssymb}
\usepackage{mathrsfs}  

\usepackage{graphicx}
\usepackage{float}

\usepackage[normalem]{ulem}

\newcommand{\avg}[1]{\langle{#1}\rangle}

\title{Analog Hawking radiation from a spin-sonic horizon in a two-component Bose-Einstein condensate}

\author{\firstname{Anna} \lastname{Berti} \CDRorcid{0000-0003-3073-9554}\IsCorresp}
\address{Pitaevskii BEC Center, INO-CNR and Dipartimento di Fisica, University of Trento (Italy)}
\email[A. Berti]{anna.berti@ino.cnr.it}

\author{\firstname{Lennart} \lastname{Fernandes} \CDRorcid{0000-0003-4994-6051}}
\address{Center for Quantum Phenomena, Department of Physics, New York University (USA)}
\address{Theory of Quantum and Complex Systems, University of Antwerp (Belgium)}

\author{\firstname{Salvatore} \lastname{Butera} \CDRorcid{0000-0002-5980-3938}}
\address{School of Physics and Astronomy, University of Glasgow, Glasgow G12 8QQ (UK)}

\author{\firstname{Alessio} \lastname{Recati} \CDRorcid{0000-0002-8682-2034}}
\addressSameAs{1}{Pitaevskii BEC Center, INO-CNR and Dipartimento di Fisica, University of Trento (Italy)}

\author{\firstname{Michiel} \lastname{Wouters} \CDRorcid{0000-0003-1988-4718}}
\addressSameAs{3}{Theory of Quantum and Complex Systems, University of Antwerp (Belgium)}

\author{\firstname{Iacopo} \lastname{Carusotto} \CDRorcid{0000-0002-9838-0149}}
\addressSameAs{1}{Pitaevskii BEC Center, INO-CNR and Dipartimento di Fisica, University of Trento (Italy)}

\begin{abstract} 
We theoretically study stimulated and spontaneous Hawking emission from an analog horizon for spin modes in a two-component Bose-Einstein condensate, both with and without a coherent coupling between the two components. We highlight the conceptual and practical advantages that these systems offer to the experimental observation of the phenomenon, namely the massive nature of elementary excitations and the experimental accessibility of the different quadratures of the spin excitations. In particular, we go beyond the relativistic regimes previously addressed in the literature, and identify various observables that show a signature of the Hawking process, as well as additional features associated with the massive nature of the modes, such as \textit{undulations}. Semi-analytical calculations of the scattering properties of the horizon and of two-point correlation functions of the emitted radiation in an ideal stationary setup are supported by time-dependent numerical simulations based on Gross-Pitaevskii and Bogoliubov theory.
\end{abstract}

\begin{document}
\maketitle

\section{Introduction}
First predicted by Hawking in 1974 \cite{hawking1974black, hawking1975particle}, the thermal emission from astrophysical black holes is one of the most surprising and fascinating consequences of quantum effects in curved spacetime. 
According to Hawking's prediction, black holes of mass $M$ are expected to radiate as black bodies 
%with temperature
%\begin{equation}
 %   k_B T_H = \frac{\hbar c^3}{8\pi G M}
 %   \label{eq:Hawkingtemp}
%\end{equation}
as a result of the interplay between quantum fluctuations and the curvature of spacetime in the vicinity of a gravitational singularity. 
In the absence of a counteracting accretion mechanism, the Hawking emission is anticipated to lead to the evaporation of the black hole \cite{hawking1974black}. However, the Hawking emission has so far escaped direct observation in the astrophysical context, as the Hawking temperature of a typical black hole mass is orders of magnitude smaller than the cosmic microwave background temperature ($\sim 2.7 K$) \cite{fixsenTemperatureCosmicMicrowave2009}, and its evaporation time longer than the age of the Universe \cite{frolovBlackHolePhysics1998}.

Nevertheless, the purely kinematic nature of the Hawking effect \cite{visser1998hawking} offers the possibility of constructing analogies between quantum field theory in curved spacetime and condensed matter systems \cite{barcelo2011analogue}, allowing to observe this phenomenon in experimentally accessible platforms such as classical fluids \cite{rousseaux2008observation, rousseaux2010horizon, torres2017rotational}, polaritons \cite{nguyen2015acoustic, gerace2012analog}, optical systems \cite{philbin2008fiber, belgiorno2010hawking} and ultracold gases \cite{lahav2010realization}. 
In the context of quantum fluids, e.g. Bose-Einstein condensates (BEC), analog black holes are generated by engineering a steadily-flowing stationary state that features an accelerating velocity profile. Where the flow velocity $v$ exceeds the speed of sound $c$, a sonic horizon separating the regions of supersonic and subsonic fluid flow acts as a point-of-no-return for long-wavelength phononic excitations, similarly to the event horizon which marks the boundary between the interior and exterior of a black hole.

Within the hydrodynamic regime, the analog Hawking temperature of a sonic horizon in a BEC is related to the spatial derivative of the speed of sound and flow velocity at the location of the horizon \cite{visser1998hawking, macher2009black}
%\begin{equation}
%    k_B T_H = \frac{\hbar }{2\pi } \frac{\partial(c-v)}{\partial x}, 
%    \label{eq:analogTH}
%\end{equation}
and is often lower than the physical temperature of ultracold atomic gases, preventing a direct observation of the phenomenon in realistic experimental setups. To circumvent this limitation, a classical analog of the Hawking effect can be 
observed in the scattering of wave packets on the sonic horizon. Despite being a stimulated variant of the Hawking process, the scattering properties of the horizon carry relevant information regarding its spontaneous counterpart, such as the greybody factor of the analog black hole and its Hawking temperature. 
Moreover, a characteristic signature of the occurrence of spontaneous emission, sometimes referred to as the \textit{Hawking moustache}, is found in the two-point correlations between density fluctuations propagating on opposite sides of the horizon \cite{recati2009bogoliubov, carusotto2008numerical, larre2012quantum}. This observable is robust against finite temperature effects \cite{carusotto2008numerical} and has been exploited to claim the first experimental observation of the Hawking effect in a single component condensate \cite{steinhauer2016observation, munoz2019observation, kolobov2021observation}. 
So far, the Hawking emission has escaped observation in any other analog system. 

This paper is devoted to the theoretical study of the Hawking process for spin modes in a two-component Bose-Einstein condensate \cite{abad2013study, larre2013hawking}.
In relation to previous studies \cite{larre2013hawking, syu2022analogous, syu2023analogous}, we aim to explicitly show how the availability of two channels of elementary excitations results in both conceptual and practical advantages to the experimental study of the Hawking effect with respect to a single-component fluid.
Our purpose in this regard is two-fold. Firstly, our focus on spin modes in a mixture allows us to conceptually extend the analysis to massive excitations by introducing a resonant coherent coupling between the two components. Differently from Refs. \cite{syu2022analogous, syu2023analogous}, we do not restrict our analysis to small values of the coherent coupling.
As all our calculations are based on the Gross-Pitaevskii and zero-temperature Bogoliubov theories, we are not restricted to the hydrodynamic regime and are instead able to consider parameters' regimes beyond the gravitational analogy but closer to typical experimental conditions \cite{cominotti2022observation}. As we will see, the notion of an analog horizon can be extended to parameters' ranges that do not allow for the definition of a sound velocity, yet the features characterizing the Hawking emission are preserved in such non-relativistic regime.
Secondly, the experimental accessibility of additional degrees of freedom, such as the relative phase between the two components, opens the possibility of identifying observables in which the intensity of the Hawking signal overcomes by orders of magnitude that of the typical moustache observed in density correlations, thus potentially facilitating its measurement. 

%\ar{It would be nice to emphasize the main results...we have a complete description of the system: reflection and transmission in a peculiar geoemtry + correlation functions + dynamics of the black hole formation... I will propose something}

The paper is structured as follows: after reviewing the main properties of binary BEC mixtures in Sec.\ref{sec:bogo}, we introduce the idealized one-dimensional spin-sonic black hole in Sec.\ref{sec:spinsonic} and propose a realistic implementation of the same setup in Sec.\ref{sec:realistic}. The following sections are devoted to stimulated Hawking emission: the scattering properties of the horizon, anticipated in Sec.\ref{sec:scattering}, are probed with time-dependent GP simulations in Sec.\ref{sec:GPtimedep} and used to estimate the greybody factor and Hawking temperature of the analog black hole in Sec.\ref{sec:THgreybody}. The last part of the manuscript focuses on the analysis of the spontaneous Hawking emission through correlation functions: semi-analytical results and time-dependent numerical simulations, both based on Bogoliubov theory, are reported in Sec.\ref{sec:corr} and Sec.\ref{sec:Bogotimedep}, respectively. Moreover, a conceptually interesting configuration which exploits the properties of the system in the vicinity of a critical point is examined in Sec.\ref{sec:critical}.
Lastly, in Sec.\ref{sec:concl} we draw conclusions and briefly discuss future perspectives. 

In the Appendices we briefly review and summarize the theoretical and numerical techniques we used to derive the results presented in this manuscript and we discuss the effects of a smooth horizon on the Hawking signal.

%%%%%%%%%%%%%%%%%%%%%%%%%%%%%%%%%%%%%%%%%%%%
\section{Coherently coupled mixtures and their collective excitations}
%%%%%%%%%%%%%%%%%%%%%%%%%%%%%%%%%%%%%%%%%%%%
\label{sec:bogo}

We consider a spinor Bose gas composed of atoms of mass $m$, subject to an external potential $V(x,t)$, and condensed in two different internal states $\ket\uparrow, \ket{\downarrow}$. 
Atoms interact with intra- and inter-component interaction strengths $g, g_{\uparrow\downarrow}$ and the two components are coherently coupled by an external coupling field of Rabi frequency $\Omega>0$, which we assume to be resonant with the atomic transition.
The mean field dynamics of this system is governed by two Gross-Pitaevskii (GP) equations ~\cite{pitaevskii2016bose, abad2013study} for the order parameters $\psi_{\uparrow, \downarrow}(x,t)$:
\begin{align}
    &i\hbar\frac{\partial\psi_\uparrow}{\partial t} = -\frac{\hbar^2\nabla^2}{2m} \psi_\uparrow + \Big(V+gn_\uparrow+g_{\uparrow\downarrow} n_{\downarrow} \Big)\psi_\uparrow-\frac{\hbar\Omega}{2} \psi_{\downarrow} \notag\\
    &i\hbar\frac{\partial\psi_\downarrow}{\partial t} = -\frac{\hbar^2\nabla^2}{2m} \psi_\downarrow + \Big(V+gn_\downarrow+g_{\uparrow\downarrow} n_{\uparrow} \Big)\psi_\downarrow-\frac{\hbar\Omega}{2} \psi_{\uparrow}
    \label{Eq:GPE}
\end{align}
where $n_{\uparrow, \downarrow} = |\psi_{\uparrow, \downarrow}|^2$ are the two atomic densities. 
For uniform mixtures of total density $n = n_\uparrow+n_\downarrow$, as long as $g_{\uparrow\downarrow}< g + \hbar\Omega/n$, the ground state is characterized by identical order parameters in the two components $\psi_\uparrow=\psi_\downarrow$, i.e. corresponds to a $\mathbb{Z}_2$-symmetric configuration. 

According to the Bogoliubov theory briefly summarized in Appendix \ref{sec:App_A}, small perturbations of the ground state then belong to either one of two independent branches of collective excitations, associated to perturbations of the total density $n$ and of the relative (spin) density $n_\uparrow-n_\downarrow$: hence, the two channels are usually referred to as density, hereafter identified by the subscript 0, and spin branches. 
The frequency $\omega$ and momentum $k$ of the excitation modes are related through the Bogoliubov dispersion relations \cite{abad2013study}: 
\begin{align}
    \omega^{(0)}_\pm(k) &= \pm \sqrt{\frac{\hbar k^2}{2m} \left( \frac{\hbar k^2}{2m} + \frac{2\mu_0}{\hbar} \right)}, \label{eq:Bogodendisp}\\
    \omega_\pm(k) &= \pm \sqrt{\left(\frac{\hbar k^2}{2m}+ \Omega \right)\left( \frac{\hbar k^2}{2m} + \frac{\kappa n}{\hbar} + \Omega \right)}, \label{eq:Bogodisp}
\end{align}
with $2\mu_0 = (g+g_{\uparrow\downarrow})n$ the density interaction energy and $\kappa = g-g_{\uparrow\downarrow}$ the interaction constant for spin modes. %The former describes perturbations of the total density, while the latter is associated to excitations of the relative (spin) density.
The presence of a coherent coupling does not affect the density branch, which is always phononic at low momenta $k < 1/\xi_0$, with speed of sound $c_0 = \sqrt{\mu_{0}/m}$ and healing length $\xi_0 = \hbar/mc_0$. On the other hand, it renders spin modes effectively massive by opening a gap of size 
\begin{equation}
    \Delta = \sqrt{\Omega (\Omega + \kappa n/\hbar)}
    \label{eq:dispgap}
\end{equation}
in the dispersion relation. If $|\kappa|n/\hbar \ll \Delta \sim \Omega$, spin modes behave as free particles with an almost parabolic gapped dispersion. 

The phononic nature of spin modes is only recovered in the opposite limit $\hbar\Delta \ll |\kappa|n$, that is, either $\Omega \sim 0$ (with $\kappa > 0$) or $\hbar\Omega + \kappa n \sim 0$ (with $\kappa < 0$). In both cases the sound speed for spin modes is $c \simeq \sqrt{|\kappa|n/2m}$ and the dispersion is linear at low enough momenta $k < 1/\xi$, where $\xi=\hbar/mc$ is the healing length for spin modes. 
In the absence of a coherent coupling, $\Omega = 0$, the atom number in each component is individually conserved, so a Goldstone mode appears in the spin channel as a consequence of a U(1) symmetry for the relative phase between the two components: exciting arbitrary relative phase perturbations is costless. The condition $\hbar\Omega + \kappa n = 0$, instead, defines the critical point for a para-to-ferromagnetic quantum phase transition \cite{cominotti2023ferromagnetism}: the closing gap in this case is due to the spontaneous breaking of the $\mathbb Z_2$ symmetry for the exchange of the two components and relative density excitations become extremely soft. 

Rigorously speaking, the gravitational analogy \cite{barcelo2011analogue} can be applied to two-component BECs only if spin modes are phononic, namely either if the coupling is small:
\begin{equation}
    \sqrt{\frac{\hbar\Omega}{\kappa n+\hbar\Omega}} \ll 1 \qquad (\kappa > 0),
    \label{eq:KGmassive1}
\end{equation}
or in the proximity of the critical point for the para-to-ferromagnetic phase transition:
\begin{equation}
    \sqrt{\frac{\hbar\Omega+\kappa n}{\hbar\Omega}} \ll 1 \qquad (\kappa < 0).
    \label{eq:KGmassive2}
\end{equation}
For this reason, throughout this paper we will refer to these two parameter regimes as \textit{relativistic}, whereas the opposite limit $|\kappa| n \ll \hbar\Omega$ is identified as \textit{non-relativistic}. 
%Notice that, while the dispersion relation \eqref{eq:Bogodisp} is identical in the two regimes \eqref{eq:KGmassive1} and \eqref{eq:KGmassive2}, the physical properties of the mixture are entirely different, due to the different sign of $\kappa$. Such difference is appreciable when considering the static structure factor for spin modes:
%\begin{equation}
%    \mathcal S(k) = \sqrt{\frac{\hbar^2k^2/2m + \hbar\Omega}{\hbar^2k^2/2m + \hbar\Omega +\kappa n}}
%    \label{eq:struct}
%\end{equation}
%which determines the response of the mixture to perturbations: $\mathcal S(0) \ll 1$ in the small coupling regime \eqref{eq:KGmassive1}, whereas $\mathcal S(0) \gg 1$ in the vicinity of the critical point \eqref{eq:KGmassive2}. As expected, in the latter case the system is extremely sensitive to (relative) density perturbations.

%\paragraph{The spin vector formalism}
By analogy with two-level systems, the state of a two-component condensate can be represented on a Bloch sphere through a spin vector $\mathbf S$ \cite{cominotti2023ferromagnetism} with components:
\begin{equation}
\begin{split}
    S_1 &= 2 \Re(\psi_\uparrow^*\psi_\downarrow),\\
    S_2 &= 2\Im(\psi_\uparrow^*\psi_\downarrow),\\
    S_3 &= n_\uparrow-n_\downarrow.
\end{split}
    \label{eq:spinvector}
\end{equation} 
For a $\mathbb{Z}_2$-symmetric mixture in the ground state, the spin vector reads $\mathbf S = (n, 0, 0)$. Assuming small perturbations of the densities of the two components $\delta n_\uparrow, \delta n_\downarrow$ and of the relative phase $\delta\phi$, the spin vector is modified as $\mathbf S = \big( n + \delta n_\uparrow+\delta n_\downarrow, n\delta\phi, \delta n_\uparrow-\delta n_\downarrow \big)$. Hence, to linear order, the second and third components of $\mathbf S$ are sensitive to spin modes, while the first component is perturbed by total density excitations. In particular, $S_2$ and $S_3$ are proportional to the relative phase and relative density perturbations and can be seen as the two quadratures of the spin excitation modes.

\section{The spin-sonic black hole}
\label{sec:spinsonic}
To study analog Hawking radiation in a Bose-Einstein condensate, one requires a fluid flow that mimics a black hole geometry.
For a single-component system, the simplest configuration with this property involves a uniform one-dimensional system steadily flowing with velocity $v$ \cite{carusotto2008numerical, recati2009bogoliubov, larre2012quantum}. 

In a simplest configuration, a step-like modulation of the interaction constant
\begin{equation}
    g(x) = g_{u}\Theta(-x) + g_{d}\Theta(x), 
    \label{eq:interactionstep}
\end{equation}
where $\Theta$ is the Heaviside step function, separates the upstream region (for $x<0$, labelled $u$), where the flow is subsonic, that is $c_u= \sqrt{g_un/m} > v$, from the downstream region (for $x>0$, labelled $d$), which is instead supersonic, $c_d = \sqrt{g_d n/m} < v$. The surface at $x=0$ thus plays the role of an analog sonic horizon. Even though the discontinuous interaction profile \eqref{eq:interactionstep} breaks the hydrodynamic approximation and suggests %while Eq.\eqref{eq:analogTH} would predict 
an infinite Hawking temperature, 
explicit calculations shows that the emission temperature $T_H$ remains finite and of the order of $\mu/k_B$, where $\mu$ is the chemical potential of the system.

Configurations involving a smoother horizon or inhomogeneous density profiles, which reproduce more closely the experimental conditions \cite{steinhauer2016observation}, have also been considered in the theoretical literature (see, for instance, \cite{larre2012quantum, macher2009black}): the general features of the Hawking signal are qualitatively identical to the simplest discontinuous case \eqref{eq:interactionstep} and the Hawking temperature recovers the prediction of the hydrodynamical approximation.
% In particular, the emission temperature $T_H$ is always finite and of the order of $\mu/k_B$, where $\mu$ is the chemical potential of the system. 
%For the sake of simplicity, we will only analyze the analytically tractable case of a step modulation of the interaction constants in a uniform system.

The aim of this work is to generalize this configuration to the spin channel of a binary mixture with $n_\uparrow=n_\downarrow=n/2$, such that the $\mathbb Z_2$ symmetry of the system implies a complete decoupling between the dynamics of excitations of the total and relative density. We thus consider a uniform mixture of total density $n$ and assume that a spin-sonic horizon is generated by a step modulation of 
the spin interaction constant,
\begin{equation}
\kappa(x) = \kappa_u \Theta(-x) + \kappa_d \Theta(x). \label{eq:mu_modulation}
\end{equation}
Conversely, the density interaction constant $(g+g_{\uparrow\downarrow})$ and the corresponding energy $\mu_0$ are kept uniform in space with a density speed of sound $c_0>v$ such that no sonic horizon is seen by density modes.
A similar configuration was analyzed in \cite{larre2013hawking}. In addition to the spatial modulation of the spin interaction constant, we also assume that the two components are coupled by a spatially inhomogeneous Rabi coupling:
\begin{equation}
    \Omega(x) = \Omega_u \Theta(-x) + \Omega_d \Theta(x). 
    \label{eq:omega_modulation}
\end{equation}
According to Eq. \eqref{Eq:GPE}, there exists a $\mathbb{Z}_2$-symmetric stationary state of the mixture of the form 
\begin{equation}
    \psi_\uparrow(x,t)=\psi_\downarrow(x,t) = \sqrt{\frac n 2} \,e^{iqx}e^{-i\mu t/\hbar}, \label{eq:hawking_stationarystate}
\end{equation}
where $q=mv/\hbar$ is the momentum associated with the fluid flow and
\begin{equation}
    \mu = \frac{1}{2} m v^2 + \mu_0 + V(x) - \frac{\hbar\Omega(x)}{2}. 
\end{equation}
Notice that if $\Omega_u\ne \Omega_d$, the jump at $x=0$ in the Rabi frequency has to be compensated by an external potential, $V(x) = \hbar\Omega(x)/2$, in order to guarantee the stationarity of the state \eqref{eq:hawking_stationarystate}.  

As already pointed out, in the presence of a coherent coupling, the dispersion relation for spin modes in a comoving reference frame \eqref{eq:Bogodisp} is gapped and a definition of the speed of sound is only possible in a limited range of parameters' values, that is, in the regimes of validity of the gravitational analogy \eqref{eq:KGmassive1}, \eqref{eq:KGmassive2}. 
In any other case, the role of the speed of sound is played by the Landau critical velocity, defined as
\begin{equation}
	c_{d,u} \equiv \min_{k} \frac{\omega_+(k, x\gtrless 0 )}{|k|}
 \label{eq:landaucritspeed}
\end{equation}
In order to generate an analog horizon for spin modes, it is thus necessary to fix $\kappa(x)$ \eqref{eq:mu_modulation} and $\Omega(x)$ \eqref{eq:omega_modulation} so that the Landau critical velocity is larger than the flow velocity in the upstream region and smaller than the flow velocity in the downstream region, i.e., $c_d < v < c_u$.

\begin{figure}[ht]
	\centering
	\includegraphics[width = 0.6\linewidth]{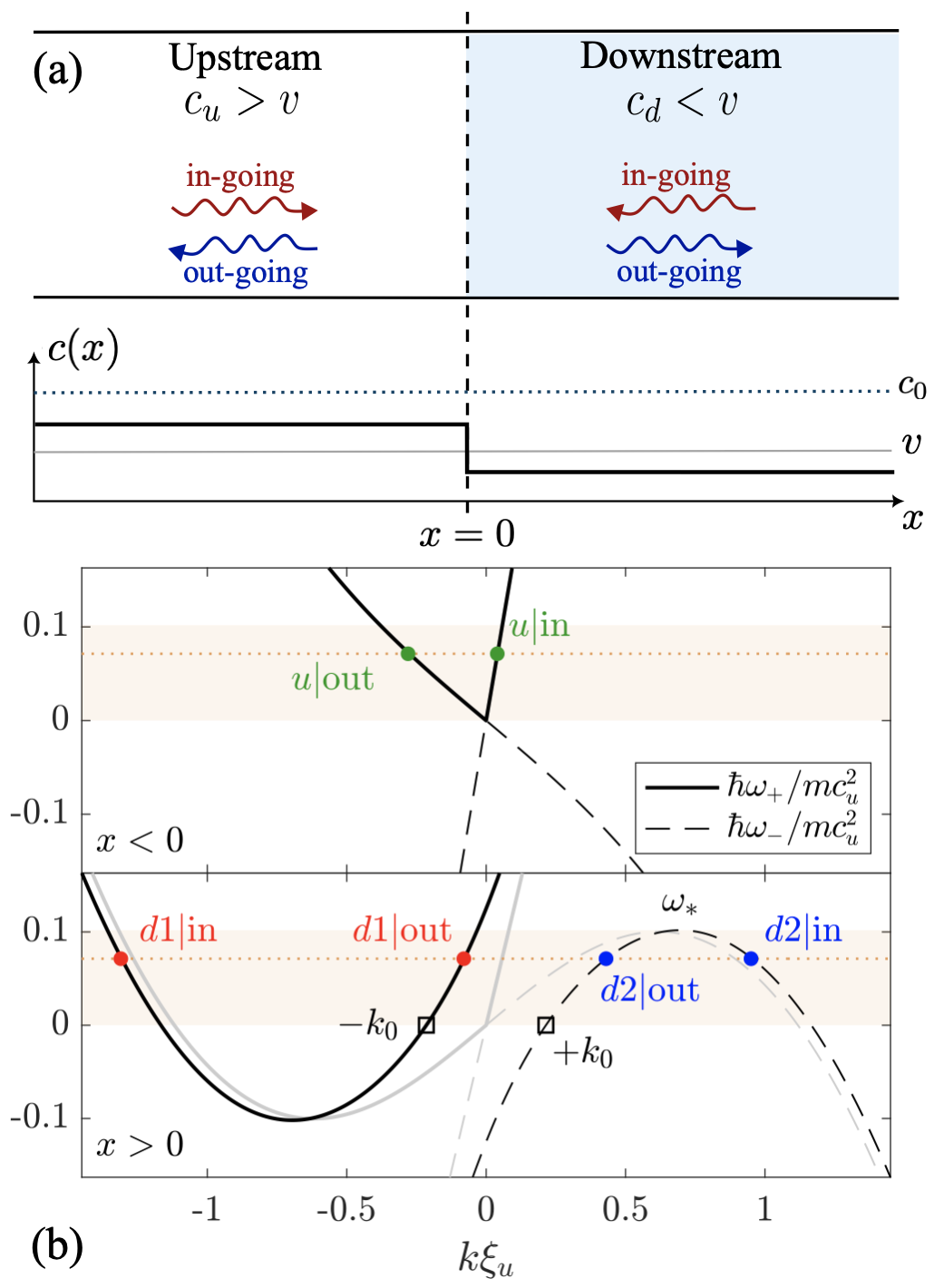}
    %\tikz \draw (0,0) rectangle (0.65\linewidth, 0.37\linewidth);
	\caption{(a) Schematic of the analog black hole toy-model configuration considered in this work. The analog horizon, located at $x=0$, separates the upstream ($u$) region where the Landau critical speed for spin modes $c_u$ exceeds the flow velocity $v$, from the downstream ($d$) region, where instead the opposite holds $v > c_d$. This step configuration can be realized while keeping the speed of density-sound $c_0=\sqrt{\mu_0/m}$ uniform and larger than $v$, so that the density channel is everywhere subsonic. The red (blue) arrows in the upper panel indicate the group velocity direction for in-going (out-going) modes. (b) Typical example of spin dispersion relation on the two sides of the analog horizon with (black curves) and without (grey curves) coherent coupling in the downstream region. The colored dots indicate all the propagating modes which exist at a generic frequency $\omega < \omega_*$ (dotted line), while the empty squares identify zero-frequency modes, which have a non-zero wavevector if $\Omega_d\ne 0$. The shaded area represents the frequency range $[0, \omega_*]$ in which negative energy modes are available. }
	\label{fig:sketch}
\end{figure}

The condition $c_d < v$ is crucial to the appearance of a sonic horizon: negative (positive) norm modes in the downstream region, once Doppler-shifted to account for the flow velocity, acquire positive (negative) frequency and thus carry negative energy in the lab frame. The existence of these solutions is the key ingredient enabling the onset of Hawking physics, as pairs of particles with opposite energy can be emitted on either side of the horizon while conserving the total energy of the system.  
A schematic drawing of the resulting spin-sonic black hole configuration is shown in Fig.~\ref{fig:sketch}, together with examples of the dispersion relation on the two sides of the horizon.

Analog Hawking radiation of massive particles in a spin-sonic black hole configuration with homogeneous coherent coupling ($\Omega_u = \Omega_d\ne 0$) has been characterized in Refs. \cite{syu2022analogous, syu2023analogous}: in this case the condition for the presence of a horizon, $c_u > v > c_d$, constrains the Rabi frequency to a limited range of extremely small values. In order to explore different regimes, we instead consider a coherent coupling with different intensity on the two sides of the horizon, $\Omega_u \ne \Omega_d$. Since the presence of a gap in the upstream region reduces the intensity of Hawking radiation emitted in the black hole exterior \cite{syu2022analogous}, we set $\Omega_u=0$ throughout this work. 
The absence of a gap in the upstream region requires the spin interaction energy $\kappa_u n/2$ to be positive, and allows us to define a speed of sound $c_u = \sqrt{\kappa_u n/2m}$ and a healing length $\xi_u = \hbar/mc_u$ for spin modes propagating in the black hole exterior. Their frequency is given by a Doppler-shifted gapless Bogoliubov dispersion:
\begin{equation}
    \omega_u^\pm(k) = %vk + \sqrt{ \frac{\hbar k^2}{2m}  \left( \frac{\hbar k^2}{2m} + \frac{\kappa_u n}{\hbar} \right) } = 
     k v \pm  |k|c_u \sqrt{  1+ \frac{\xi_u^2 k^2}{4}  }. \label{eq:disp_up_labframe}
\end{equation}
The condition $c_u > v$ guarantees that propagating modes with real wave-vector and positive norm necessarily have positive frequency.

On the contrary, let us assume that the coherent coupling is not necessarily vanishing in the downstream region.
From the perspective of the gravitational analogy, varying the intensity of the coherent coupling in the downstream region $\Omega_d$ keeping $\Omega_u=0$ amounts to changing the nature of the analog black hole (in particular its greybody factor, as we will see later on), without affecting the spacetime geometry of the black hole exterior.   

For generic values of $\Omega_d$, the role of the speed of sound is replaced by the Landau critical velocity. A direct calculation from \eqref{eq:landaucritspeed} leads to:
\begin{equation}
    mc_d^2 = \kappa_d n/2 + \hbar\Omega_d + \hbar\Delta_d
\end{equation}
Notice that $c_d$ coincides with the speed of sound $\simeq \sqrt{|\kappa_d|n/2m}$ if the gap $\Delta_d$, defined as in \eqref{eq:dispgap}, is small or vanishing, that is, in the two relativistic regimes \eqref{eq:KGmassive1} and \eqref{eq:KGmassive2}, while, in the opposite limit $\hbar\Omega_d \gg |\kappa_d| n$ of a strong Rabi coupling the gravitational analogy breaks and we find the density-independent value $c_d \sim \sqrt{2\hbar\Omega_d/m}$. 

The supersonic condition $c_d < v$ is fulfilled by choosing the parameters in the downstream region such that:
\begin{equation}
    1-\frac{\kappa_d n}{2mv^2} > \sqrt{\frac{2\hbar\Omega_d}{mv^2}}
\end{equation}
As already pointed out, this requirement guarantees the existence of a range of positive frequencies $(0<\omega<\omega_*)$ for which both positive and negative norm modes exist in the downstream region, with dispersion given by:
\begin{align}
    \omega_{d}^\pm(k) = k v \pm  \sqrt{\left( \frac{\hbar k^2}{2m} +\Omega_d \right) \left( \frac{\hbar k^2}{2m} + \Omega_d + \frac{\kappa_d n}{\hbar} \right) } \label{eq:disp_down_labframe}
\end{align}
The threshold frequency is $\omega_* = \max(\omega_d^-)$.

The simplifying assumptions of uniform density and infinitely steep horizon enable us to perform simple semi-analytical calculations. These are based on the evaluation of the scattering properties of the horizon via direct calculation of the corresponding scattering matrix~\cite{recati2009bogoliubov, carusotto2008numerical}. Such results are supported and further extended by time-dependent GP and Bogoliubov simulations, which allow us to consider more general and realistic configurations, to investigate the effects of a smooth sonic horizon with finite width $\sigma_x$ and to determine how the Hawking signatures develop after the analog horizon is created.  

%\anna{Thanks to the uniform density and the infinitely steep horizon, the model we just introduced allows us to perform simple semi-analytical calculations. These results are supported by time-dependent GP and Bogoliubov simulations: the discretization of space required by the numerical analysis inevitably prevents us from investigating horizons' widths smaller than the grid spacing; in order to avoid numerical issues, the step profiles \eqref{eq:mu_modulation}, \eqref{eq:omega_modulation} are thus replaced by a smooth sigmoid function with the same asymptotic values at $x \to \pm \infty$. More details about the numerical simulations are reported in} \lf{Appendices \ref{appendix:GPsim} and ????}.

\section{Realistic implementation}
\label{sec:realistic}
While in the following of this work we focus mostly on the idealized step configuration considered in the previous Section or on its straightforward generalization to a smooth horizon, it is instructive to briefly discuss a more realistic implementation of the spin-sonic black hole. 

As a first point, the toy model of Fig.~\ref{fig:sketch} is one-dimensional (1D). While Bose-Einstein condensation is, strictly speaking, not possible in one spatial dimension, finite size effects allow to work with quasi-1D cigar-shaped condensates in which the transverse degrees of freedom are frozen \cite{gorlitzRealizationBoseEinsteinCondensates2001, lahav2010realization, steinhauer2016observation}. 
In the case of two-component BECs, the spin degree of freedom can be effectively 1D without the need of reaching this one-dimensional regime. For most atomic species, contact interaction constants are comparable in strength, $g\sim g_{\uparrow\downarrow}$, resulting in a speed of spin-sound much lower than that of density-sound, $c \ll c_0$. Consequently, the healing lengths satisfy $\xi \gg \xi_0$. If the transverse size $R$ of a cigar-shaped condensate is chosen such that $\xi_0 < R < \xi$, the density profile is well approximated by the Thomas-Fermi (TF) solution, but the dynamics of spin modes remains frozen in the transverse direction and is thus effectively one-dimensional. Such a configuration has been realized, for instance, in Ref. \cite{cominotti2022observation}.

Another feature of the idealized configuration is the infinitely steep variation of the speed of sound $c$ across the horizon.
As already pointed out, the non-hydrodynamic nature of the step does not affect the relevant properties of Hawking emission, such as its thermal character. However, the horizon width influences the value of the Hawking temperature and therefore the intensity of the signal in two-point correlations: the steeper the step, the stronger the \textit{moustache} (see Appendix \ref{appendix:GPsim} for a detailed discussion).
Experimentally, the steepness of the sonic horizon is bound by the (density) healing length $\xi_0$ and is often of the order of several $\xi_0$ \cite{steinhauer2016observation}. 
For two-component BECs, such width can still be much smaller than the healing length for spin modes $\xi$, effectively rendering the horizon infinitely steep and maximizing the Hawking temperature. 

\begin{figure}[tbp]
    \includegraphics[width = 0.6\linewidth]{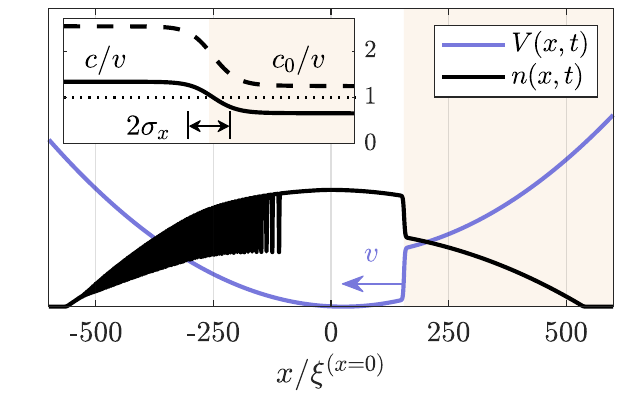}
    \caption{Experimentally realizable spin-sonic horizon in a harmonically trapped condensate. A scanning step potential $V(x,t)  \sim \tanh[(x - vt) / \sigma_x] + 1$ induces a modulation of the atomic density $n(x,t)$, dividing the condensate into regions of spin-subsonic (left) and spin-supersonic flow (right, shaded). The sudden appearance of a sharp boundary generates a shock wave, seen propagating away from the horizon in the subsonic region. The inset shows a closeup of the velocity profile around the horizon of width $\sigma_x=2\xi^{(x=0)}$.}
    \label{fig:horizon_geometry}
\end{figure}

For spin modes, in the toy model we consider, the horizon is generated by a step in the spin interaction constant $\kappa$ while the total density is kept uniform; this modulation is not readily implemented in experiments. Alternatively, a non-uniform interaction energy $\kappa n /2$ can be achieved by modulating the density via an external potential. This approach was employed in the first realization of sonic black hole in a single-component BEC \cite{lahav2010realization} and subsequent observations of spontaneous analog Hawking radiation \cite{steinhauer2016observation,munoz2019observation, kolobov2021observation}. Figure~\ref{fig:horizon_geometry} shows a spin-sonic horizon generated with a similar setup: a stationary, harmonically trapped two-component fluid is scanned from right to left by a modulated potential at velocity $v$. 
The step-like potential causes a buildup of atoms in the upstream region, conversely decreasing the density in its wake downstream. 
As shown in the inset, the speed and magnitude of the modulated potential can be tuned to generate a sonic horizon for spin modes while keeping the density channel subsonic throughout the gas, making the configuration robust against density perturbations. 

%--------------------
%--------------------

%\anna{PARAMETERS: Steinhauer's paper: $c_u/v_u\sim 2.4$, $c_d/v_d\sim 0.25$, $n_u\xi_u \sim 60$, $n_d\xi_d \sim 30$, $L/\xi_u\sim 80$, $T_H \sim 1.2 nK$. Faraday paper: $n\xi \sim 7000, L/\xi \sim 150$, $\kappa n \sim 2\pi\hbar\times 300 Hz$, $\Omega \sim 2\pi \hbar\times 30Hz$. This value of the density would give a Hawking temperature of $\sim 5 nK$ with the velocities of Steinhauer's paper.  }

The observation of an analog Hawking effect for spin excitations in this setup is still facing a few several experimental issues.
Firstly, as is the case for a scalar Bose gas \cite{kamchatnovGenerationDispersiveShock2012}, the introduction of the scanning potential may generate an undesired shock wave in the density of the fluid, propagating away from the horizon as shown in Fig.~\ref{fig:horizon_geometry}. Its eventual reflection at the edge of the condensate introduces a finite lifetime for the undisturbed horizon.
Secondly, not all parameter regimes discussed in Sec. \ref{sec:bogo} are straightforwardly accessible to state-of-the-art experiments.
Given the necessity of working with low densities in order to increase the visibility of quantum effects \cite{steinhauer2016observation}, and due to the relatively small value of $\kappa$ with respect to the average interaction constant, accessing the relativistic massive regime defined by \eqref{eq:KGmassive1} and considered in Refs.\cite{syu2022analogous, syu2023analogous} would require to control the Rabi frequency at least at the level of a few Hz. 
Approaching the critical point for the para-to-ferromagnetic phase transition \eqref{eq:KGmassive2} presents an even greater challenge, as it requires, in addition to a fine tuning of the Rabi frequency, a change of sign in $\kappa$ across the horizon, a configuration which is only achievable by exploiting spatially modulated Feshbach resonances~\cite{chienSpatiallyVaryingInteractions2012}.
For this reason, we focus in the remainder of this work solely on the standard gapless case considered in Ref.\cite{larre2013hawking} and on the non-relativistic regime $|\kappa_d| n \sim \hbar\Omega_d$. 
Finally, the low interaction energy $\kappa n/2$ results in a correspondingly low temperature of spontaneously emitted Hawking radiation which, already in the case of a scalar Bose gas, is notoriously weak in comparison to undesired fluctuations due to the thermal cloud of the condensate, three-body heating and shot noise encountered in measurements \cite{macher2009black, wuster2008phonon}. 

The last issue is typically solved by looking at the correlation functions.
In particular, in contrast to a single-component Bose gas, a binary mixture provides experimental access to all the quadratures of the spin excitation modes by looking at the different components of the spin vector \eqref{eq:spinvector}. For instance, $S_2$, which is closely related to the relative phase between the two components, can be measured by applying a $\pi/2$-pulse before measuring the relative density of the mixture, that is, $S_3$. The availability of additional degrees of freedom thus makes it possible to observe signatures of the Hawking effect in several observables and, as we are going to see in the next Sections, take advantage of the stronger Hawking signal in some of them.

\section{The scattering matrix formalism}
\label{sec:scattering}

As a first step in the study of analog Hawking physics in BEC mixtures, we analyze the scattering properties of Bogoliubov waves on the spin-sonic horizon. The scattering matrix formalism has been first developed to characterize single component systems \cite{recati2009bogoliubov, carusotto2008numerical}, and then extended to treat two-component mixtures in the absence of coherent coupling \cite{larre2013hawking} or in the relativistic regime $\hbar\Omega \ll \kappa n$ \cite{syu2022analogous, syu2023analogous}. However, since it does not rely on the gravitational analogy but exclusively on the validity of Bogoliubov theory, it can be exploited in all ranges of parameters.

For the analytically tractable case of a piecewise uniform fluid of total density $n$, the field perturbations are easily written as a sum of plane waves within each uniform region. Given the stationarity of our configuration, it is convenient to work in frequency rather than momentum space. For each frequency $\omega>0$ in the laboratory frame, there exist four solutions to the equation $\omega_{u,d}^\pm(k) = \omega$ in both the subsonic and supersonic regions. In the upstream region, we find two positive norm modes with real $k$ (green dots in Fig.~\ref{fig:sketch}, labelled $u$), in addition to two negative norm modes with complex conjugate momenta $\Re (k)\pm i \Im (k)$ (not shown): of these, one is exponentially growing for $x\to -\infty$, and thus represents an unphysical solution, while the other is evanescent in the same limit and contributes to the perturbation only close to the horizon. 
The same holds in the downstream region if $\omega > \omega_*$; if instead $\omega < \omega_*$ four propagating solutions with real momentum exist, two with positive norm (red dots in Figs.\,\ref{fig:sketch}, labelled $d1$) and two with negative norm (blue dots in Figs.\,\ref{fig:sketch}, labelled $d2$). Among all the real-$k$ modes, some move towards the sonic horizon (in-going, labelled $in$), while others propagate away from it (out-going, labelled $out$), depending on the sign of their group velocity $w_k \equiv \partial \omega/\partial k$.

Bogoliubov eigenstates of frequency $\omega>0$ are built as a linear combination of plane waves, each one associated to one of the available propagating modes at that frequency. As it is briefly summarized in Appendix \ref{appendix:scattering}, the relative amplitudes of in-going and out-going modes, which we indicate with $\alpha, \beta$ respectively, are related through a \textit{scattering matrix} $\mathcal M(\omega)$ of dimension $D\times D$, where $D$ is the number of in-going and out-going modes available at frequency $\omega$. In the frequency range $\omega < \omega_*$, propagating modes are available in all three branches and the scattering solution reads
\begin{equation}
    \begin{bmatrix} \beta_u \\ \beta_{d1} \\ \beta_{d2} \end{bmatrix} = \mathcal M(\omega) \begin{bmatrix} \alpha_u \\ \alpha_{d1} \\ \alpha_{d2} \end{bmatrix},
    \label{eq:scattering}
\end{equation}
whereas for $\omega > \omega_*$ (or if neither region is supersonic), we find
\begin{equation}
    \begin{bmatrix} \beta_u \\ \beta_{d1} \end{bmatrix} = \mathcal M(\omega) \begin{bmatrix} \alpha_u \\ \alpha_{d1}  \end{bmatrix}.
\end{equation}
The coefficient $\mathcal M_{rr'}(\omega)$ describes the scattering of the $r'$-th in-going mode onto the $r$-th out-going one. The square modulus $|\mathcal M_{rr'}(\omega)|^2$ represents therefore the reflectivity/transmittivity of the in-going mode $r'$ into the out-going mode $r$.
In order to guarantee energy conservation, $\mathcal M$ must satisfy the normalization condition
\begin{equation}
\mathcal M^\dagger \eta \mathcal M = \eta = \mathcal M \eta \mathcal M^\dagger,
\label{eq:scatteringnorm}
\end{equation}
where $\eta$ is a diagonal square matrix of dimension $D$ whose diagonal elements coincide with the norm $\sigma_r = \pm 1$ of the corresponding mode, that is $\eta_{rr'} = \sigma_r \delta_{rr'}$. 

If the $d2$ branch is not available, $\eta = \text{diag}(1,1)$ coincides with the identity matrix and $\mathcal M$ is unitary. The problem then reduces to a standard scattering problem in which only positive energy modes are involved; each incoming packet is partially transmitted and partially reflected, and the transmittivity and reflectivity sum up to 1:
\begin{equation}
    |\mathcal M_{ur}|^2 + |\mathcal M_{d1r}|^2 = 1, \qquad r = u, d1.
\end{equation}
The unitarity of the scattering matrix also proves the equivalence of the basis of in-going and out-going modes in the decomposition of the field perturbation.  

If instead propagating $d2$ modes are available, that is, for $\omega < \omega_*$, energy conservation is satisfied with $\eta = \text{diag}(1, 1, -1)$ and results in
\begin{equation}
    |\mathcal M_{ur}|^2+|\mathcal M_{d1r}|^2 - |\mathcal M_{d2r}|^2 = \sigma_r,
    \label{eq:encons}
\end{equation}
for all the modes $r=u, d1, d2$.
As a consequence of the minus sign in \eqref{eq:encons}, reflectivities and transmittivities are not bound to be smaller than 1. Also, the scattering matrix is pseudo-unitary rather than unitary, signalling the inequivalence between the representation of the perturbation in terms of in-going and out-going modes. This mathematical fact is at the basis of spontaneous particle creation in quantum field theories on curved spacetimes.  

%Similarly, if a gap of size $\omega_\text{min}$ is present in the subsonic region, the $u$-branch is not available for $\omega < \omega_\text{min}$. Hence, the scattering matrix has again dimension $2\times 2$ but $\eta = \text{diag}(1,-1)$. In this frequency range, the horizon behaves as a perfect mirror: any in-going mode in the downstream region is completely reflected back, but split into two packets with opposite energy. By energy conservation during the scattering process, the original wavepacket must then be amplified: a clear signature of such superradiant processes is found if the reflection coefficients $|\mathcal S_{d1d1}|^2, |\mathcal S_{d2d2}|^2$ exceed unity for $\omega < \omega_\text{min}$. \anna{Has this something to do with Andreev reflection?}

Upon quantization of the modes, the Bogoliubov theory allows to study spontaneous Hawking emission by characterizing the radiation flowing from the interior (downstream region) to the exterior (upstream region) of the black hole \cite{recati2009bogoliubov, larre2013hawking}. 
The standard quantization procedure, briefly summarized in Sec.\ref{appendix:quantization}, consists in replacing the complex amplitudes $\alpha_r,\beta_r$ of in-going and out-going scattering modes with creation/annihilation operators $\hat a_r, \hat b_r$ satisfying the commutation relations:
\begin{equation}
    [\hat a_r, \hat a_{r'}^\dagger] = [\hat b_r, \hat b_{r'}^\dagger] = \sigma_r \delta_{rr'}
\end{equation}
Notice how, for the negative norm $d2$ modes, creation and annihilation operators are exchanged\footnote{The calculations shown in this Section can be equivalently performed by explicitly exchanging creation and annihilation operators for the $d2$ modes, in order to recover canonical commutation relations for all the modes. This is often done in previous literature, see for instance Refs. \cite{recati2009bogoliubov, larre2012quantum}. }. This fact is at the origin of Hawking emission: within our formalism, the annihilation operators $\hat a, \hat b$ for in-going and out-going modes are related through the scattering matrix as in \eqref{eq:scattering}:
\begin{equation}
    \hat b_r(\omega) = \sum_{r'} \mathcal M_{rr'}(\omega) \hat a_{r'}(\omega)
    \label{eq:scattering1}
\end{equation}
This relation allows to compute expectation values for out-going modes starting from those of the in-going ones: as long as the $d2$ mode is available, these two quantities do not coincide, signalling that a process of spontaneous particle creation is taking place. 
In particular, assuming the system is initially at zero temperature, a finite intensity of out-going particles in the upstream region can be detected, proportionally to the $|\mathcal M_{ud2}(\omega)|^2$ coefficient  %Hence, this element of the scattering matrix is referred to as the Hawking spectrum of the analog black hole.
that 
can be thought of as the (zero-temperature) Hawking emission spectrum \cite{recati2009bogoliubov, carusotto2008numerical,larre2012quantum}.
.%, since it is associated to the number of $u$ particles that are spontaneously emitted by the analog horizon in the black hole exterior, even in the absence of any in-going mode.

Within this picture, the transmittivity of $d$ modes into the $u$ mode (or, equivalently, the reflectivity of the horizon with respect to the $u$ mode), 
\begin{equation}
        \Gamma(\omega) = |\mathcal M_{d1u}(\omega)|^2 - |\mathcal M_{d2u}(\omega)|^2 = 1 - |\mathcal M_{uu}(\omega)|^2
    \label{eq:greybody}
\end{equation}
can be interpreted as a greybody factor of the analog black-hole.

%%%%%%%%%%%%%%%%%%%%%%%%%%%%%%%%%%%%%%%%%%%%
\section{Probing the spin-sonic horizon}
\label{sec:GPtimedep}
%%%%%%%%%%%%%%%%%%%%%%%%%%%%%%%%%%%%%%%%%%%%
With the spin-sonic horizon described in terms of a scattering matrix, its spectral properties can be probed through the scattering of plane waves on the horizon in a classical field description as briefly summarized in Appendix \ref{appendix:GPsim}.
% Indeed, the spectrum \eqref{eq:Bogodisp} of Bogoliubov excitations underlying the Hawking effect equally arises for small amplitude oscillations on a steady state solution of the GP equations \eqref{Eq:GPE} \cite{pitaevskii2016bose}. Stimulated excitation of these modes therefore enables one to infer the elements of the scattering matrix, which in turn carry information about the spontaneous Hawking effect, in itself an inherently quantum mechanical phenomenon.

\begin{figure}[tb]
    \includegraphics[width = .6\linewidth]{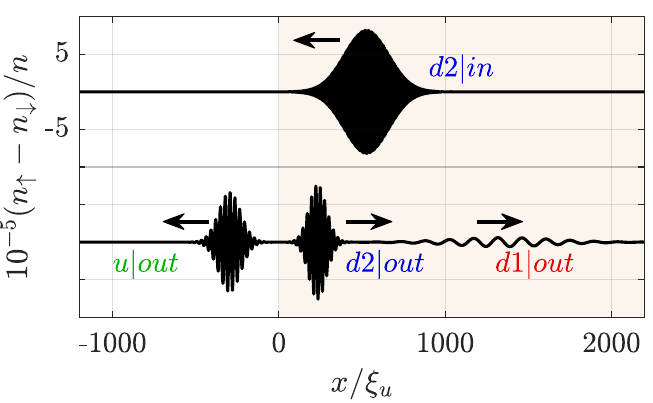}
    \caption{Illustration of an in-going $d2$ wave-packet of frequency $\omega/\omega_*=0.12$ scattering on a sharp spin-sonic horizon for $\kappa_d/\kappa_u=0.25$ and $\Omega_d=0$ and generating out-going propagating wave-packets in all three branches $u, d1, d2$. The shaded area identifies the downstream region. }
    \label{fig:Bragg_example}
\end{figure}

% In particular, the coefficient $|\mathcal M_{ud2}(\omega)|^2$ can be thought of as the (zero-temperature) Hawking emission spectrum \cite{recati2009bogoliubov, carusotto2008numerical}, since it is associated to the number of $u$ particles that are spontaneously emitted by the analog horizon in the black hole exterior, even in the absence of any in-going mode. 
% Moreover, the transmittivity of $d$ modes into the $u$ mode (or, equivalently, the reflectivity of the horizon with respect to the $u$ mode), 
% \begin{equation}
%         \Gamma(\omega) = |\mathcal M_{d1u}(\omega)|^2 - |\mathcal M_{d2u}(\omega)|^2 = 1 - |\mathcal M_{uu}(\omega)|^2
%     \label{eq:greybody}
% \end{equation}
% can be interpreted as a greybody factor of the analog black-hole.
Spin modes may be excited without disturbing the density degree of freedom by a periodic modulation of the transverse harmonic confinement~\cite{cominotti2022observation} or by applying a spin-selective optical pulse \cite{carusotto2006bragg, farolfi2020observation, kim2020observation}. In the latter approach, the application of a laser beam on a small region of the condensate \cite{kim2020observation} results in the generation of a localized wave packet, which may propagate and scatter on the sonic horizon.
Upon collision with the horizon, an in-going wave scatters into all out-going modes available at the excited frequency $\omega$.
For $\omega>\omega_*$, a single excitation is emitted on either side of the horizon, along the $u|\text{out}$ and $d1|\text{out}$ branches of the dispersion, respectively. Below the threshold $\omega < \omega_*$, however, an additional $d2|\text{out}$ signal characteristic of the Hawking effect is excited downstream, as shown in Fig.~\ref{fig:Bragg_example}.
The spin dispersion on both sides of the sonic horizon, as well as the frequency dependence of all the scattering matrix coefficients, can be reconstructed by extracting the wavevector and the relative amplitudes of scattered wave packets as a function of $\omega$ (not shown). 

For the ideal scenario of a sharp horizon shown in Fig.~\ref{fig:sketch}, the elements of the scattering matrix can be computed through a semi-analytical calculation based on Bogoliubov theory (see Appendix \ref{appendix:scattering}). 
However, this is no longer possible for a spin-sonic horizon of finite width $\sigma_x$ such as the one encountered in the realistic setup sketched in Fig.~\ref{fig:horizon_geometry}, where one has to resort to a complete simulation of the GP or Bogoliubov equations. In the following, we will explicitly compare semi-analytic predictions obtained for the infinitely steep profile with GP results for a horizon of non-negligible width, in order to highlight the differences. The effect of the horizon's spatial profile on signatures of the Hawking process is discussed in more detail in Appendix \ref{appendix:horizoneffect}, where we show that the agreement between the two approaches is excellent in the limit $\sigma_x \to 0$.

%--------------------
%--------------------

\begin{figure}[tb]
    \includegraphics[width = \linewidth]{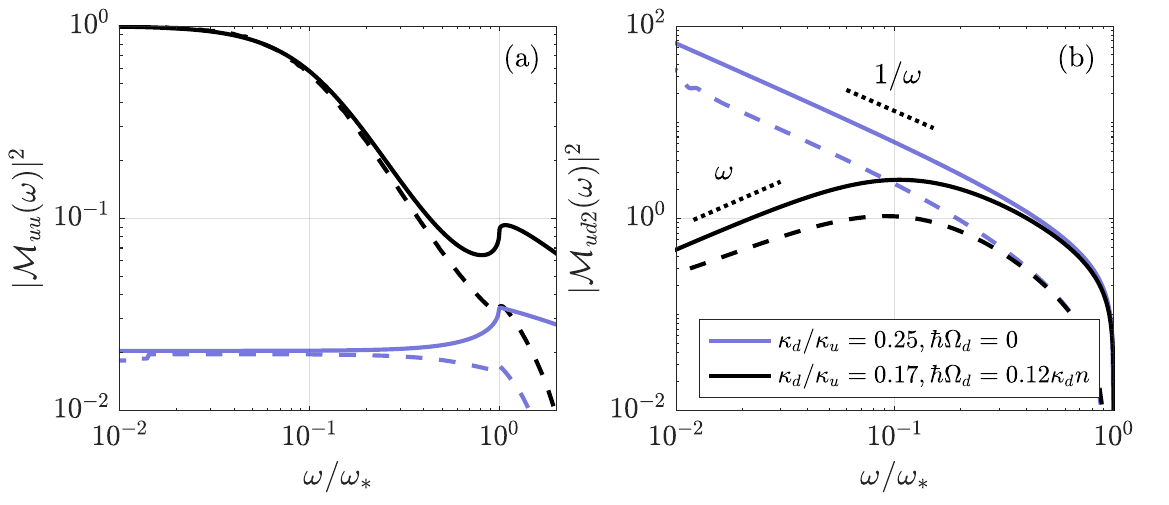}
    %\caption{Frequency dependence of the Hawking spectrum $|\mathcal M_{ud2}|^2$ for $\kappa_d/\kappa_u = 0.25$, $\Omega_d = 0$ (blue lines) and $\kappa_d/\kappa_u = 0.17$, $\hbar\Omega_d = 0.12 \kappa_d n$ (\iac{solid black line}); parameters in the upstream region are $c_u/v = 4/3$ and $\Omega_u=0$ in both cases. The dispersion relations for these two \iac{sets of parameters are} reported in Fig.~\ref{fig:sketch}. Solid lines indicate the Bogoliubov semi-analytic prediction for an infinitely steep profile, while dashed lines represent the respective transmission coefficients inferred through wave packet scattering on a horizon of finite width $\sigma_x=\sqrt{2}\xi_u$.}
    %\label{fig:S_omega_ideal}
    \caption{Frequency dependence of the scattering coefficients $|\mathcal M_{uu}|^2$ (a) and $|\mathcal M_{ud2}|^2$ (b) for $\kappa_d/\kappa_u = 0.25$, $\Omega_d = 0$ (blue lines) and $\kappa_d/\kappa_u = 0.17$, $\hbar\Omega_d = 0.12 \kappa_d n$ (black lines); parameters in the upstream region are $c_u/v = 4/3$ and $\Omega_u=0$ in both cases. The dispersion relations for these two sets of parameters are reported in Fig.~\ref{fig:sketch}. Solid lines indicate the Bogoliubov semi-analytic prediction for an infinitely steep profile, while dashed lines represent the respective transmission coefficients inferred through wave packet scattering on a horizon of finite width $\sigma_x=\sqrt{2}\xi_u$.}
    \label{fig:S_omega_ideal}
\end{figure}

As a general fact, we observe that, in contrast to the astrophysical case, the sonic horizon does not act as a perfect black body: indeed the reflectivity of the $u$ mode  $\abs{\mathcal{M}_{uu}(\omega)}^2$, shown in Fig.~\ref{fig:S_omega_ideal}(a) is non-vanishing. In other words, the Hawking emission spectrum $|\mathcal M_{ud2}(\omega)|^2$ does not coincide with a perfect black-body thermal spectrum, but features a (possibly frequency-dependent) greybody factor \eqref{eq:greybody}. 
As expected for astrophysical non-rotating black holes, no superradiant amplification of the in-going $u$ mode is observed, as $|\mathcal M_{uu}(\omega)|^2 < 1$ for all frequencies. Superradiant amplification is instead observed for downstream modes $d1,d2$ inside the analog black hole; for instance $|\mathcal M_{d2d2}(\omega)|^2>1$ for all frequencies, by definition [see Eq. \eqref{eq:encons}]. This phenomenon has no counterpart in the gravitational context, since the propagation of particles in the black hole interior towards the event horizon is not allowed, as it would require a superluminal dispersion relation.

Observable consequences of the Hawking effect are encoded in the scattering coefficients of in-going $d$ modes, impinging on the horizon from the black hole interior.
Figure~\ref{fig:S_omega_ideal}(b) shows, for instance, the frequency dependence of the spontaneous Hawking spectrum $|\mathcal M_{ud2}(\omega)|^2$ in the absence (blue) and presence (black) of a coherent coupling; dashed lines refer to the scattering matrix coefficient inferred through time-dependent GP simulations of a smooth horizon, while solid lines indicate the semi-analytical Bogoliubov prediction of the idealized scenario \eqref{eq:mu_modulation}. 
The approaches agree qualitatively, with most notably a consistently smaller magnitude of the GP result due to the lower value of the Hawking temperature \eqref{eq:analogTH} of a horizon with a finite width $\sigma_x = \sqrt{2}\xi_u$ (see Appendix \ref{appendix:horizoneffect}).

%In the gapless case ($\Omega_d=0$, blue lines), the large amplitude of the negative norm $d2|\text{out}$ mode at low frequencies is accompanied by an amplification of the $u|\text{out}$ mode leaving the black hole, beyond the amplitude of the initial in-going mode. Energy is thus extracted from the black hole through the excitation of negative energy modes on its interior \cite{recati2009bogoliubov, larre2013hawking}. The $1/\omega$ infrared divergence of the scattering coefficients is regularized by the introduction of a finite gap in the downstream region \cite{syu2022analogous, syu2023analogous}. 
%\lf{In both the gapless ($\Omega_d=0$, blue lines) and gapped case ($\Omega_d>0$, black lines), the occurrence of a frequency range where $|\mathcal{M}_{ud2}|^2 \gg 1$ implies through the energy conservation condition \eqref{eq:encons} that a negative-norm $d2|\text{out}$ mode must be excited in the interior of the black hole, as illustrated in Fig.~\ref{fig:Bragg_example}. The pairwise excitation of $u|\text{out}$ and $d2|\text{out}$ modes, characteristic of the Hawking process, thus signifies a dissipation of energy from the black hole \cite{recati2009bogoliubov, larre2013hawking}.}

The $1/\omega$ infrared divergence of the $|\mathcal M_{ud2}(\omega)|^2$ scattering coefficient for the uncoupled mixture, clearly visible in Fig.~\ref{fig:S_omega_ideal}(b) (blue lines), is regularized in the presence of a finite gap in the downstream region \cite{syu2022analogous, syu2023analogous}.
More specifically, zero-frequency $u$ ($d$) modes cannot be transmitted in the downstream (upstream) region, thus all the transmittivities vanish in the limit $\omega\to 0$, while all the reflectivities tend to constant values \cite{coutant2012hawking}. In particular, $|\mathcal M_{uu}(\omega\to 0)|^2\to 1$ and $|\mathcal M_{ud2}(\omega\to 0)|^2\to \omega$: the Hawking spectrum is thus linear in frequency in the $\omega\to 0$ regime (see Fig.~\ref{fig:S_omega_ideal}) and reaches a maximum at a finite frequency $\overline\omega$, above which the $1/\omega$ behaviour is recovered. The dependence of $\overline{\omega}$ from the parameters is non-trivial, but our calculations show that it grows with the coherent coupling until it approaches $\omega_*$ in the critical relativistic regime \eqref{eq:KGmassive2}. 

%--------------------
%--------------------

\section{Hawking temperature and greybody factor}
\label{sec:THgreybody}

In the previous Sections we have seen that the coefficient $|\mathcal M_{ud2}(\omega)|^2$ encodes the spectrum of the Hawking emission. In order to assess its thermal nature, let us write it in the form
\begin{equation}
    |\mathcal M_{ud2}(\omega)|^2 = \frac{\Gamma(\omega)}{\exp\big[\hbar\omega/k_BT_H(\omega)]-1}, 
    \label{eq:hawkingspectrum_gapless}
\end{equation}
where $\Gamma(\omega)= 1- |\mathcal M_{uu}(\omega)|^2$ is the greybody factor defined above and we introduce the Hawking temperature parameter $T_H(\omega)$: a thermal spectrum would require a frequency-independent $T_H$, whereas a frequency-dependent $T_(\omega)$ would account for deviations from thermality.
Note that the Hawking temperature $T_H$ is unrelated to the physical temperature $T$ of the BEC under consideration: indeed all results presented in this paper are derived assuming $T=0$. The analysis of their robustness to finite temperature will be the subject of future works.

\begin{figure}[t]
    \centering
    \includegraphics[width = \linewidth]{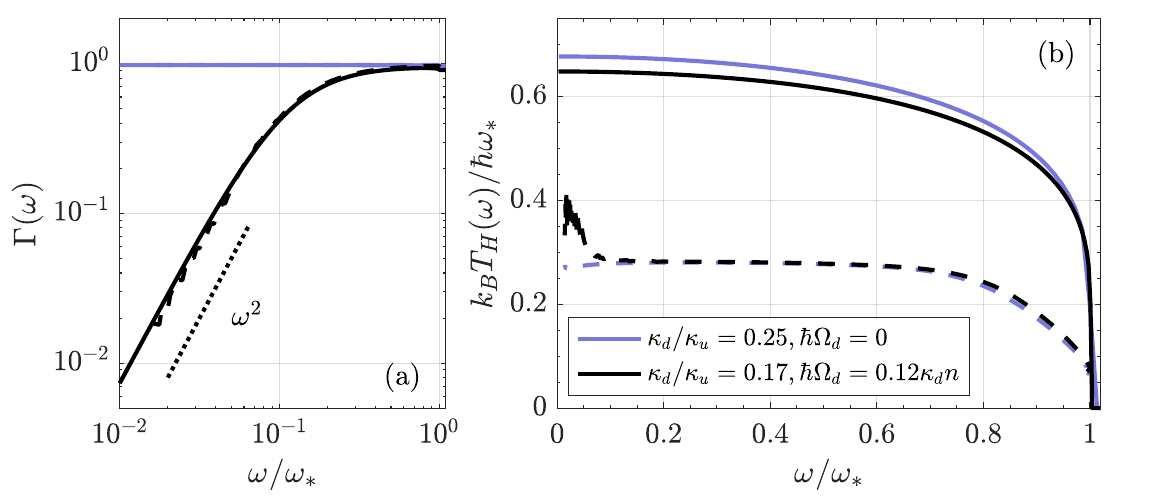}
    %\tikz \draw (0,0) rectangle (0.75\linewidth, 0.55\linewidth);
    \caption{Frequency dependence of the greybody factor (a) and the Hawking temperature (b) of the analog black hole; blue and black lines refer to $\Omega_d=0$ and $\Omega_d\ne 0$, respectively. The parameters are the same of Fig.~\ref{fig:S_omega_ideal} and were chosen so to have similar values of $\omega_* \sim 0.1 \kappa_u n/2\hbar$. The greybody factor is computed as $\Gamma(\omega) = 1-|\mathcal M_{uu}(\omega)|^2$; the Hawking temperature is then derived by inverting Eq.\,\eqref{eq:hawkingspectrum_gapless}. Full and dashed lines refer to the Bogoliubov semi-analytical prediction for a sharp horizon and to the result of GP time-dependent simulations for a smoother horizon with $\sigma_x=\sqrt{2}\xi_u$, respectively.}
    \label{fig:Hspectrum}
\end{figure}

The numerically computed results for the greybody factor and the Hawking temperature are shown in respectively the left and right panels of Fig.~\ref{fig:Hspectrum} for $\Omega_d = 0$ (blue lines) and $\Omega_d\ne 0$ (black lines).  
In the gapless case $\Omega_d=0$ \cite{recati2009bogoliubov, larre2013hawking}, given the $1/\omega$ dependence of $|\mathcal M_{ud2}|^2$, Eq.\eqref{eq:hawkingspectrum_gapless} holds with constant $T_H$ in the low-frequency limit, its value uniquely determined by the ratios $v/c_u$ and $v/c_d$ \cite{macher2009black, larre2013hawking}. For instance, the parameters of Fig.~\ref{fig:S_omega_ideal} give $k_BT_H(\omega\to 0)/\hbar\omega_* \simeq 0.68$. On the contrary, if $\omega \sim \omega_*$, $T_H$ necessarily acquires a frequency dependence: in particular, it must be vanishing for $\omega \ge \omega_*$, since $\mathcal M_{ud2}$ is identically zero if the $d2$ mode is not available. Although it could in principle be detected in experiments \cite{isoard2020departing}, this frequency-dependence is typically considered a weak deviation from thermality due to non-hydrodynamic effects, since the maximum intensity of emitted particles is found in the $\omega\to 0$ regime, where $T_H$ is approximately constant. 
The greybody factor is found to be almost constant in the whole frequency range $[0, \omega_*]$. However, as anticipated, it does not equal unity due to a residual reflectivity of the horizon with respect to the $u$ mode. With the parameters in Fig.~\ref{fig:S_omega_ideal}, we find $\Gamma(\omega\to 0) \simeq 0.98$, in excellent agreement with the analytical results of \cite{larre2013hawking}. 

In the presence of a coupling in the downstream region, $\Omega_d\ne 0$, the different behaviour of the scattering coefficients affects the greybody factor, which acquires a quadratic frequency dependence in the low-frequency regime. In the gravitational context, this behaviour has been found, for instance, for non-minimally coupled massless scalar fields in a Schwarzschild spacetime \cite{crispino2013greybody, kanti2014greybody}, where the coupling with the curvature effectively acts as a mass term in the Klein-Gordon equation.  
Remarkably, the presence of a coherent coupling in the downstream region has instead little effect on the Hawking temperature $T_H(\omega)$: the degree of thermality in the emission spectrum is preserved, despite the inaccuracy of the gravitational analogy if $\hbar\Omega_d \sim \kappa_d n$. 

While the greybody factor retrieved from GP simulations of a smooth horizon configuration is in near-perfect agreement with the semi-analytical result for a sharp horizon [left panel of Fig.~\ref{fig:Hspectrum}], the smooth horizon results in a significantly lower Hawking temperature [right panel of Fig.~\ref{fig:Hspectrum}] in accordance with the vertical offset observed in Fig.~\ref{fig:S_omega_ideal}(b). A systematic analysis of the scaling of $T_H$ with the horizon width $\sigma_x$ is included in Appendix \ref{appendix:horizoneffect}, together with an explicit discussion of the dynamical stability of the black hole configuration.

Typical experimental parameters for Sodium atoms \cite{cominotti2022observation} give a Hawking temperature of the order of a few $nK$. A physical temperature lower than $T_H$ would be necessary to directly observe Hawking phonons emitted from the analog horizon. Such small temperatures are typically not accessible in realistic experimental setups. However, it has been shown \cite{recati2009bogoliubov} that density correlations exhibit characteristic non-local features attributed to the Hawking effect which are robust against temperature effects. The analysis of such two-point correlations will therefore be the subject of the next Section.

\section{Two-point equal-time correlation functions}
\label{sec:corr}

\begin{figure*}[th]
    \centering
    \includegraphics[width=\linewidth]{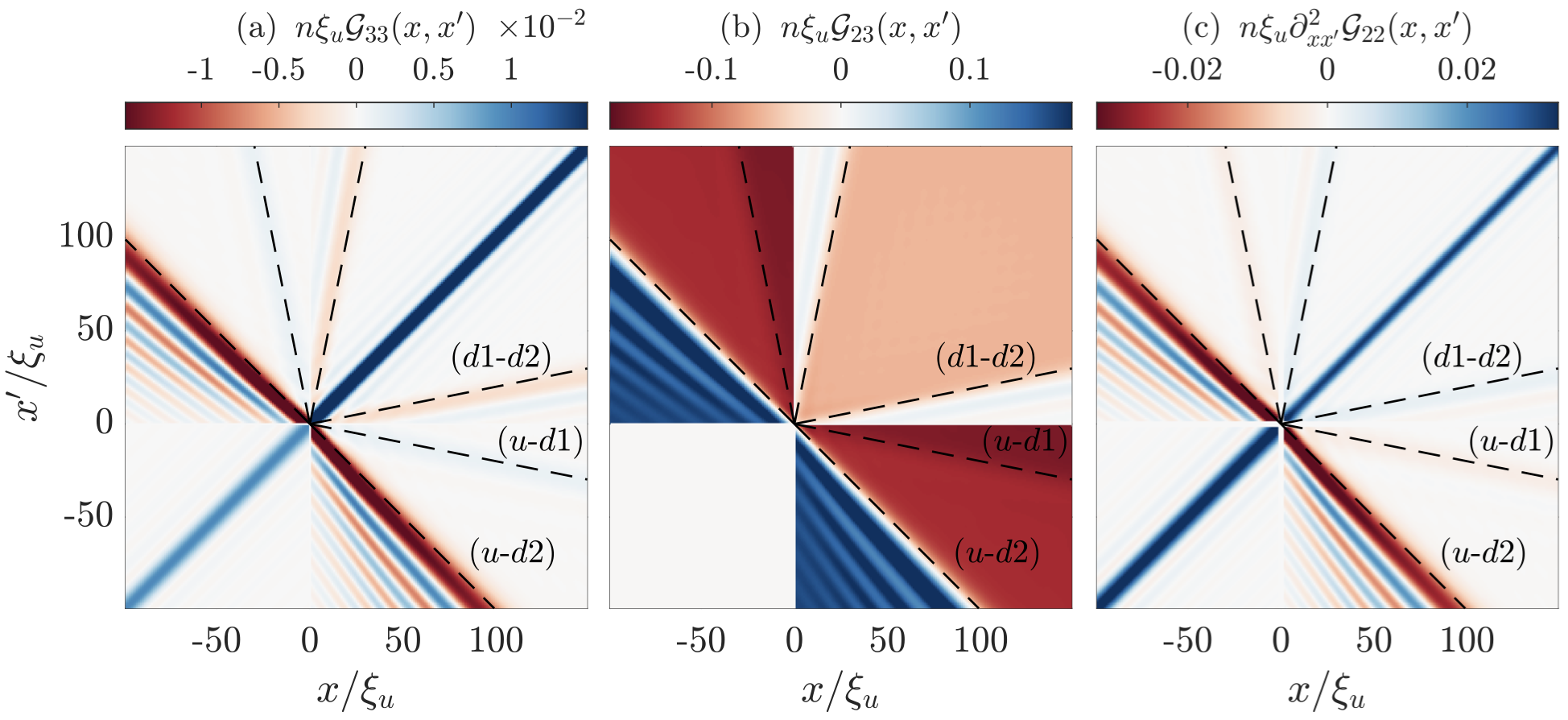}
    \caption{Equal-time correlation functions for an uncoupled two-component BEC. Parameters are identical to those of Fig.~\ref{fig:S_omega_ideal} (solid blue lines). The three panels report: (a) density correlations \eqref{eq:G33definition}; (b) density-phase correlations \eqref{eq:G23definition}; (c) the double derivative of (regularized) phase correlations, computed from \eqref{eq:G22definition} after replacing $\hat S_2(x)\to \hat S_2(x)-\hat S_2(0)$ (see discussion in the main text). Dashed black lines identify the expected location of the signal associated to each pair of modes, given by Eq. \eqref{eq:corrlines_gapless}.  }
    \label{fig:gapless}
\end{figure*}

A key observable which has been considered in previous works on single-component condensates \cite{recati2009bogoliubov, carusotto2008numerical, larre2012quantum} is the equal-time density correlation function, which, in the presence of a sonic horizon, displays characteristic non-local features associated to the spontaneous creation of pairs of particles which travel on opposite sides of the sonic horizon. In the case of a two-component condensate without coherent coupling, an analogous signal is expected in relative density correlations \cite{larre2013hawking}, that is, in the equal-time correlations of the third component of the spin vector \eqref{eq:spinvector}:
\begin{equation}
    n^2 \mathcal G_{33}(x,x') = \avg{:\hat S_3(x)\hat S_3(x'):}\,.
    \label{eq:G33definition}
\end{equation}
For simplicity, this quantity will be indicated in what follows as density correlations.

As already pointed out in Sec.\ref{sec:bogo}, however, the spin field perturbation not only affects the relative density $S_3(x)$ but also the second component of the spin vector $S_2(x)$, which is closely related to the relative phase between the two components of the mixture. It is therefore natural to define relative phase correlations, by analogy, as:
\begin{equation}
    n^2 \mathcal G_{22}(x,x') = \avg{:\hat S_2(x)\hat S_2(x'):}
    \label{eq:G22definition}
\end{equation}
Both the (relative) density correlations \eqref{eq:G33definition} and these (relative) phase correlations are the sum of a trivial contribution which is present even without a sonic horizon, and additional non-local terms which instead only contribute if the negative norm $d2$ mode is available. The latter are therefore directly associated to Hawking emission (see Appendix \ref{appenxix:corr}).  

As a last observable, let us also define a (relative) density-(relative) phase correlation function:
\begin{equation}
    n^2 \mathcal G_{23}(x,x') =\frac 1 2 \Big[ \avg{\hat S_2(x)\hat S_3(x')}+\avg{\hat S_3(x)\hat S_2(x')}\Big],
    \label{eq:G23definition}
\end{equation}
which, thanks to the symmetrized definition, is non-zero only if the $d2$ mode exists (see Appendix \ref{appenxix:corr}).

For the simplest step configuration of Fig.~\ref{fig:sketch}, correlation functions are computed semi-analytically as integrals in frequency space, since the calculation only requires the knowledge of the Bogoliubov dispersion relation, of the static structure factor and of the scattering matrix elements. More details and several useful analytical results can can be found in Appendix \ref{appenxix:corr}.
In the following, we report the results for the two cases under investigation here (see Fig.~\ref{fig:S_omega_ideal}), namely in the standard uncoupled case $\Omega_d=0$ and in the presence of a strong coherent coupling $\Omega_d$ in the downstream region only. 

\subsection{Uncoupled two-component BEC}
Let us start from the straightforward generalization of a black hole configuration for an uncoupled symmetric mixture, already considered in Ref. \cite{larre2013hawking}: as discussed in Sec. \ref{sec:bogo}, in the absence of coherent coupling, $\Omega_d=0$, the dispersion relation for spin modes is gapless both in the subsonic and supersonic region and the sonic horizon is generated by a step in the spin interaction constant $\kappa>0$, which also translates into a step-like behaviour of the speed of spin-sound $c = \sqrt{\kappa n/2m}$.

Figure~\ref{fig:gapless}(a) reports the computed two-point density correlation function \eqref{eq:G33definition}: in agreement with previous works \cite{carusotto2008numerical, recati2009bogoliubov, larre2012quantum, larre2013hawking}, it shows non-local features peaked along the straight lines (dashed black in the plot) defined by:
\begin{equation}
    x' = \frac{w_{r'}^\text{out}(0)}{w_r^\text{out}(0)} x
    \label{eq:corrlines_gapless}
\end{equation}
where $w_r^\text{out}(0)$ is the zero-frequency group velocity of the $r = u, d1, d2$ mode in the lab frame. moustaches. At zero temperature, such \textit{moustaches} are not present in the absence of the horizon and are thus directly imputed to Hawking emission: in particular, each line \eqref{eq:corrlines_gapless} is associated to a pair $(r$-$r')$ of modes, and can be interpreted as arising from the spontaneous emission of correlated particles in modes $r$ and $r'$ from the analog horizon. The strongest contribution is due to a negative correlation between the Hawking particle emitted outside the analog black hole and its partner emitted in its interior (see the $u$-$d2$ signal in Fig.~\ref{fig:gapless}). 

Even though density correlations have been a powerful experimental tool to detect Hawking emission \cite{steinhauer2016observation, munoz2019observation, kolobov2021observation}, their weak amplitude requires important experimental efforts in their measurement even when working with low atomic densities.  
Let us then fully exploit the additional degrees of freedom that a two-component BEC displays by analysing the signal appearing in other correlation functions. 
Figure~\ref{fig:gapless}(b) reports the density-phase correlation \eqref{eq:G23definition}: the result is a collection of patches in which $\mathcal G_{23}$ has different, almost constant values, separated by smooth jumps whose location is determined, once again, by the solution of \eqref{eq:corrlines_gapless}. In other words, jumps in $\mathcal G_{23}(x,x')$ correspond to peaks in $\mathcal G_{33}(x,x')$: the higher the jump in density-phase correlations, the more intense the signal in density-density correlations, i.e.,
\begin{equation}
    \mathcal G_{33}(x,x') \sim \partial_x \mathcal G_{23}(x,x'),
\end{equation}
as can be inferred from the analytical expressions reported in Appendix \ref{appenxix:corr}.
Once again, these features would be absent without an horizon; moreover, the amplitude of the signal is an order of magnitude larger than that observed in $\mathcal G_{33}$. 

In our one-dimensional configuration, due to the absence of a true condensate, the relative phase correlation $\mathcal G_{22}(x,x')$ (as well as the global phase correlation) displays an infrared divergence, formally due to the large wavelength behavior of the Goldstone mode (see Appendix \ref{appenxix:corr} for more details). As we will see in the following subsection, $\mathcal G_{22}$ is regularized by the presence of a coherent coupling, which fixes the relative phase between the two components in the ground state, explicitly breaking one of the two U(1) symmetries of the problem.

%Relative phase correlations $\mathcal G_{22}(x,x')$, instead, display an infrared divergence both for trivial configurations and for analog black holes. 
%From a physical point of view, we interpret such a divergence as due to the gauge symmetry of the system, or, equivalently, to the presence of a Goldstone mode associated to the relative phase of the two components: in each statistical sample (or experimental realization) the phase of both components is chosen randomly.  
%Indeed, as we will see in the following subsection, phase correlations are regularized by the presence of a coherent coupling, which fixes the relative phase between the two components to be $\phi=0$ in the ground state. 

In order to extract information from the phase degree of freedom, we can cure the infrared divergence by calculating $\partial^2_{xx'} \mathcal{G}_{22}$, which corresponds to the correlation of the spatial derivative of $S_2$. As shown in the Appendix \ref{appenxix:corr}, we expect such correlation to have the same structure as $\mathcal{G}_{33}$; this is confirmed by the result shown in Fig.~\ref{fig:gapless}(c).
%Alternatively, the divergence can be cured by replacing, in the definition of phase correlations \eqref{eq:G22definition}, $\hat S_2(x,t)\to \hat S_2(x,t)-\hat S_2(0, t)$, namely, by measuring $\hat S_2$ with respect to some reference point, the location $x=0$ of the sonic horizon being the most natural choice. This produces a finite correlation signal $\mathcal G_{22}(x,x') \propto xx'$, that only diverges at infinite distance from the two axis $x=0, x'=0$, where instead it vanishes, by definition (not shown). 
%Although such signal might appear featureless, the typical Hawking \textit{moustaches} are recovered by eliminating the linear dependence of $\mathcal G_{22}$ on the coordinates $x,x'$, namely by taking its first derivative with respect to both $x$ and $x'$. 
%The result is shown in Fig.~\ref{fig:gapless}(c): notice that, despite phase correlations are orders of magnitude larger than density correlations, the amplitude of the signal in Fig.~\ref{fig:gapless}(c) is comparable to the one of Fig.~\ref{fig:gapless}(a), being roughly twice as large. 
Similarly, a single derivative of $\mathcal G_{22}$ with respect to either $x$ or $x'$, followed by a symmetrization of the signal, would produce a correlation pattern similar to Fig.~\ref{fig:gapless}(b), with slightly larger amplitude (not shown).

\begin{figure}[t]
    \centering
    \includegraphics[width=0.55\linewidth]{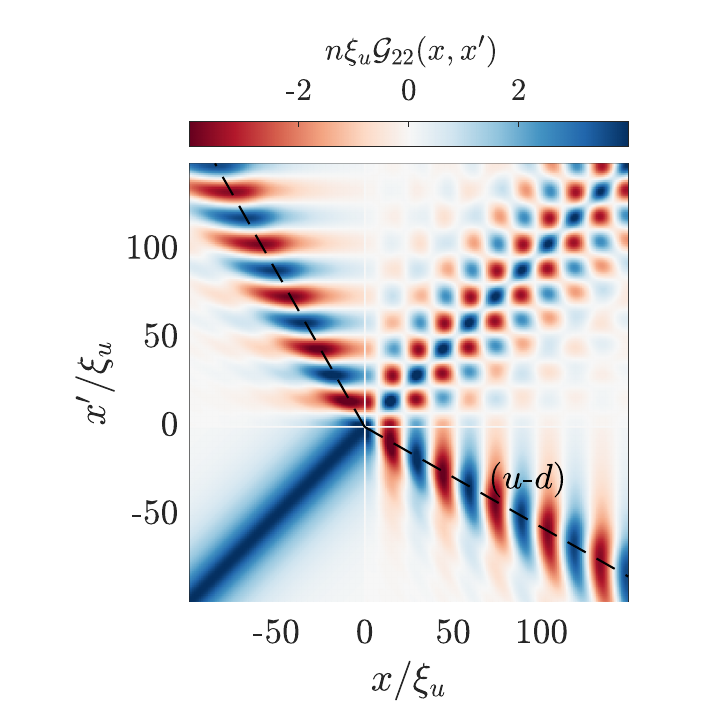}
    \caption{Phase-phase correlation \eqref{eq:G22definition} for a sharp horizon in the presence of a coherent coupling in the downstream region. Parameters are those of Fig.~\ref{fig:S_omega_ideal} (black lines), corresponding to the non-relativistic regime $\hbar\Omega \sim \kappa n$. With these parameters, the zero-frequency modes have momentum $k_0\xi_u \sim 0.2$, leading to oscillating patterns with characteristic wavelength $2\pi/k_0 \sim 30\xi_u$. Dashed black lines identify the expected location of the signal associated to $(u$-$d)$ correlations. }
    \label{fig:gapped}
\end{figure}

\subsection{Coupled two-component BEC}
While considering the phase degree of freedom might not be of great practical advantage in the experimental measurement of analog Hawking emission for an uncoupled two-component condensate, 
%our results show that phase correlations carry the same information of density correlations. 
in the presence of a coherent coupling we can directly access phase correlations. 
%and they represent a promising observable to detect Hawking emission in coherently coupled two-component condensates. 

An example is shown in Fig.~\ref{fig:gapped}, which shows phase correlations $\mathcal G_{22}(x.x')$ in the presence of a (large) coherent coupling $\Omega_d\ne 0$ in the downstream region only. In particular, a non-local correlation signal between the interior and exterior of the analog black hole 
%, which can be orders of magnitude larger than the one observed in density correlations [see Fig.~\ref{fig:gapless}(a)],
is observed in the off-diagonal $(u$-$d)$ sectors ($x\cdot x'<0$), with an amplitude that can be orders of magnitude larger than the one observed in density correlations [see Fig.~\ref{fig:gapless}(a)]. The qualitative differences with respect to the standard gapless case of Fig.~\ref{fig:gapless}(a) that are visible in Fig.~\ref{fig:gapped} stem from
the massive nature of elementary excitations in the black hole interior: the presence of zero-frequency modes with non-zero wavevector $\pm k_0$ (see Fig.~\ref{fig:sketch}) produces oscillating features known as \textit{undulations} \cite{coutant2012hawking}, with wavelength $2\pi/k_0$.  These manifest,  in the $(d$-$d)$ quadrant ($x,x'>0$), as 
%an \textit{oscillating Hawking moustache} in the $(u$-$d)$ sectors ($x\cdot x'<0$), whereas they produce 
a checkerboard pattern. A similar result has been found for two-dimensional black-hole configurations, where a finite transverse wave-vector plays the role of a mass term for longitudinal modes \cite{coutant2012hawking}. Note that checkerboard patterns associated to Hawking emission also appear in white-hole configurations \cite{mayoral2011acoustic}: in the massive case, however, the signal is not infrared divergent and its intensity is maximum along the main diagonal $x=x'$. 

Remarkably, undulations are also observed in the $(u$-$d)$ sectors $(x\cdot x' < 0)$ in the form of an \textit{oscillating Hawking mustache}, which was never found in previous works, and is the result of the correlation between massless modes emitted outside the analog black-hole with massive ones emitted in its interior.  
%\ar{however the nature of the instability and of the parttern is very different} This pattern resemble the one obtained for acoustic white-holes \cite{mayoral2011acoustic}, with the key difference that it is regularized by the presence of the coupling and thus not infrared divergent. It also shares analogy with the signal predicted for higher-dimensional black hole configurations~\cite{coutant2012hawking}\ar{remark the difference and that the oscillanting moustache has not been considered. I will call a wave guide and not higher dimension (and they got the dispersion relation wrong for a BEC...)}.
The expected location of these features is once again given by straight lines defined by \eqref{eq:corrlines_gapless}; however, since positive and negative norm modes in the downstream region have the same group velocity in the low-frequency limit, correlation signals associated to the $d1$ and $d2$ modes are superimposed. This explains why a single moustache appears in Fig.~\ref{fig:gapped}. 

Let us point out that, since we are considering a stationary configuration, our results can be considered valid at infinitely long times after the creation of the analog horizon. In a realistic experimental setup, it is only possible to observe Hawking emission at finite time $t$ after initializing the analog black hole, and this might lead to different features in the correlations at short/intermediate times after the creation of the horizon. Although an approximate, yet straightforward, estimation of $\mathcal G_{\nu\nu'}(x,x')$ ($\nu,\nu' = 2, 3$) at a time $t$ after the creation of the horizon is obtained by explicitly setting an infrared frequency cutoff, an accurate analysis of the time-dependence of correlations requires more sophisticated numerical techniques and is performed in the next Section. 

\begin{figure*}[t!]
    \centering
    \includegraphics[width=\linewidth]{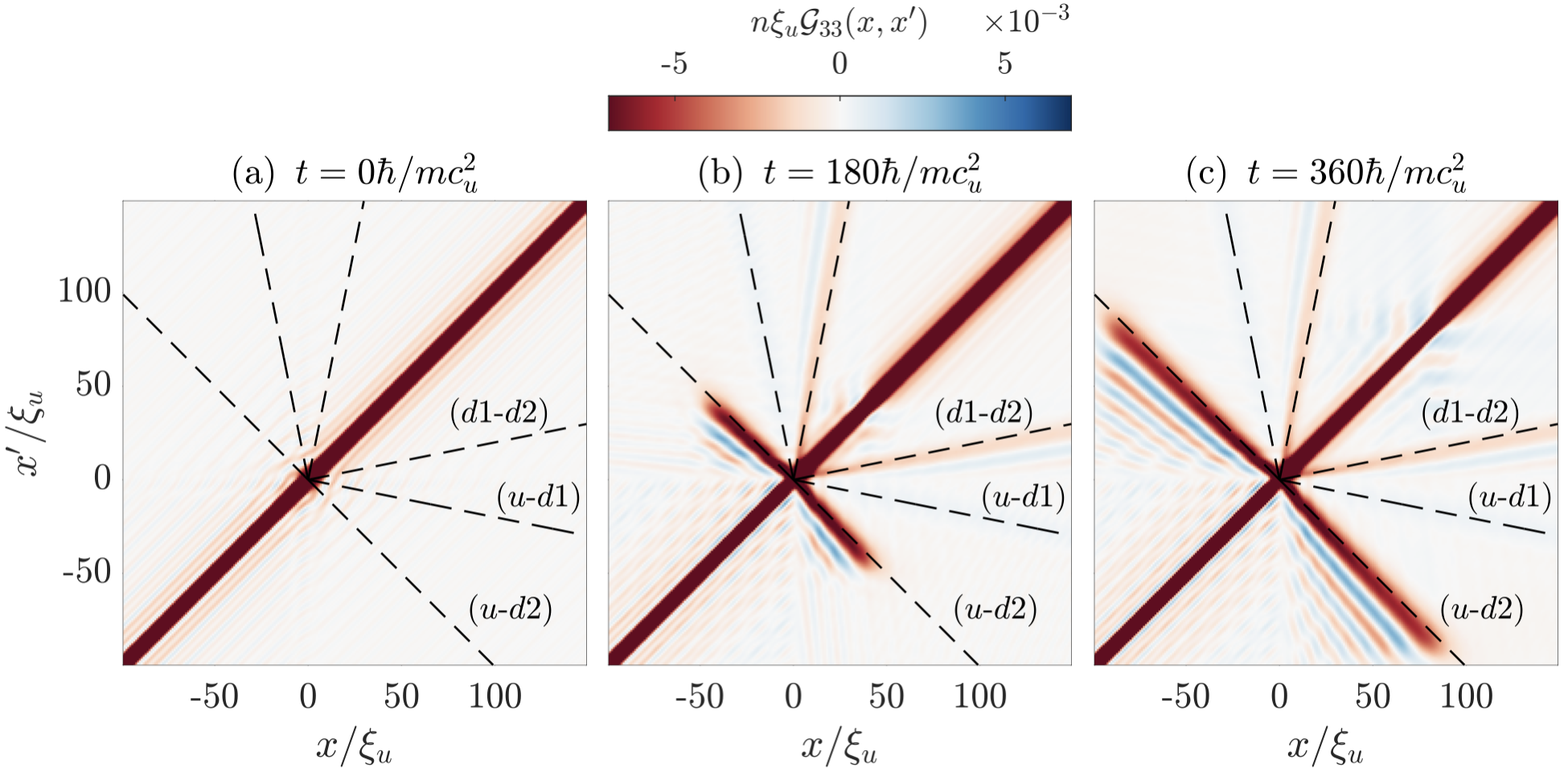}
    \caption{ Time evolution of equal-time density-density correlation functions, in an uncoupled two component condensate.%absence of coherent coupling in the downstream region. %In panels (a-c) are density-density correlations while, in panels (d-e), are density-phase correlations. 
    Three specific times are shown, that are: (a) time of formation of the horizon, $mc_u^2 t/\hbar = 0$; (b) intermediate time, $mc_u^2 t/\hbar = 180$; (c) maximum time achievable in the simulation with the setup described in the main text, $mc_u^2 t/\hbar = 360$; this is the travelling time for the white hole emission to reach the black hole due to the periodic boundary conditions implemented in the simulation (see Appendix~\ref{appendix:AppBogotimedep}). These results provide the time dependent scenario for the stationary configuration of Fig.~\ref{fig:gapless}(a), but with a smooth horizon of width $\sigma_x = 1.9\xi_u$.}
    \label{fig:gapless_time_dep_dx2}
\end{figure*}

\section{Time-dependent Bogoliubov simulations}
\label{sec:Bogotimedep}
A complete solution for the dynamics of the elementary excitations in the system is presented in this Section, based on a  numerical simulation of the full time-dependent Bogoliubov problem (see Appendix~\ref{appendix:AppBogotimedep} for details). These findings confirm and further extend the results obtained in the previous Section for the stationary configuration, by providing the complete time evolution of $\mathcal G_{\nu\nu'}(x,x')$ ($\nu,\nu' = 2, 3$), focusing in particular on the structure of the correlations shortly after the creation of the horizon. By pursuing this approach, we are also able to relax the idealised assumption of sharp, step-like horizon that was utilised in Sec.~\ref{sec:scattering} to obtain a semi-analytical solution to the scattering matrix. This permits us to investigate more realistic and experimentally relevant smooth black hole configurations, as well as to explore the effect of the horizon's size on the structure and intensity of the correlation signal (see Appendix \ref{appendix:horizoneffect}). 
%Indeed, we show that the size of the horizon not only affects the strength of the Hawking signal, because of its connection to the temperature of the emission, but it also modifies the spatial structure of the correlations in certain relevant conditions.

In our simulations we consider a one-dimensional, ring-shaped two-component BEC. Notice that the periodic boundary conditions that characterise this configuration imply the presence of both a black hole and a white hole in the system. This configuration limits the maximum time achievable in our simulations, with the limit given by the travelling time for the white hole emission to reach and interfere with the black hole. This limitation is readily overcome by increasing the length of the system up to a value that permits the investigation of the time domain of interest. We perform the numerical simulations by evolving in time the spin modes, initially obtained by diagonalising the corresponding Bogoliubov operator on top of a mean-field described by the order parameter introduced in Eq.~\eqref{eq:hawking_stationarystate}. The horizon for spin fluctuations is dynamically created by modulating both in time and in space the spin collisional interaction strength $\kappa(x,t)$ and the Rabi frequency $\Omega(x,t)$. In order to maintain the mean-field component stationary, we modify the external potential in such a way that the chemical potential of the system remains uniform in space.

\begin{figure*}[t!]
    \centering
    \includegraphics[width=\linewidth]{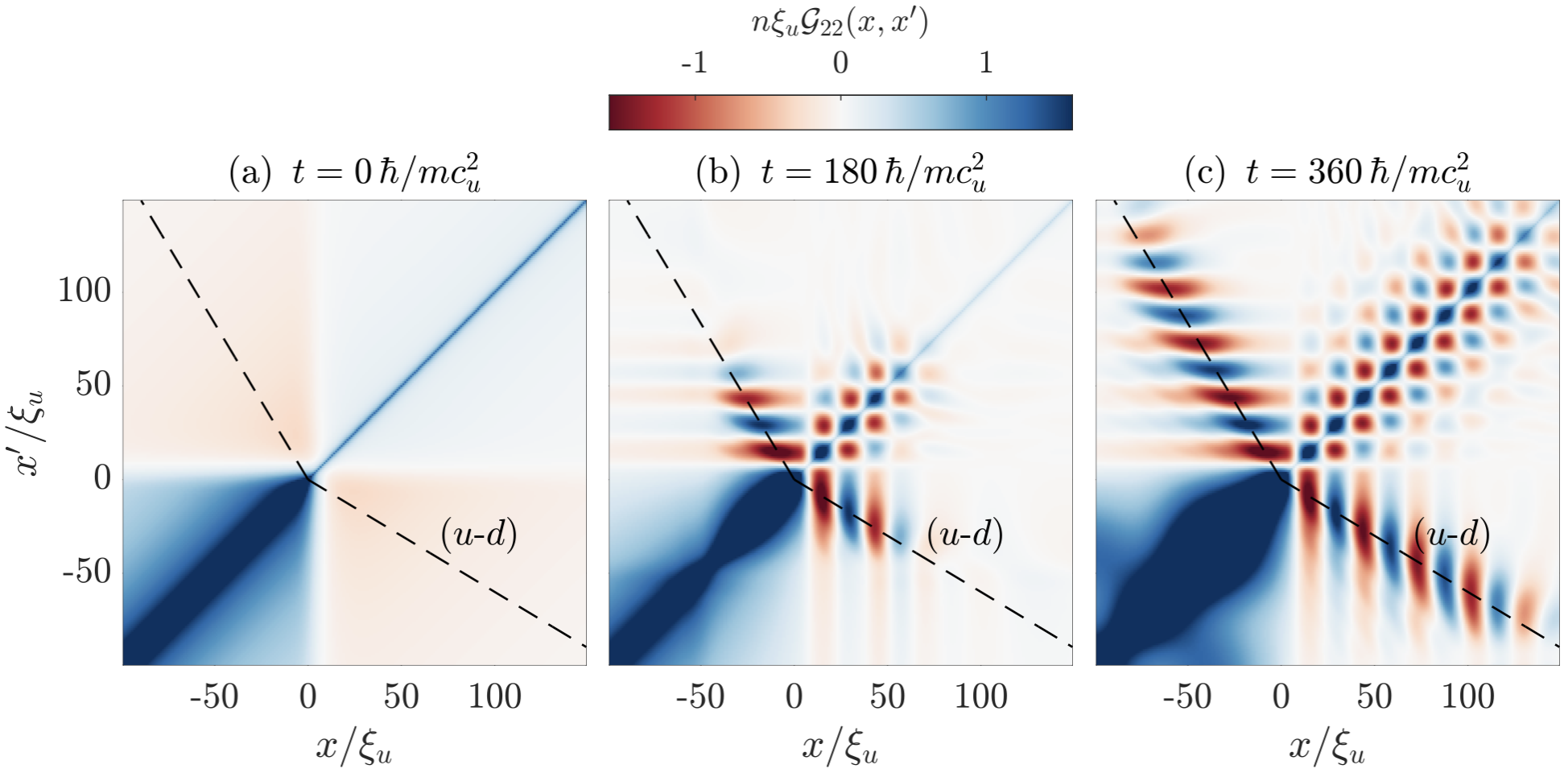}
    \caption{Time evolution of equal-time phase-phase correlation functions, in the presence of a coherent coupling in the downstream region. Panels refer to the same configuration of Fig.~\ref{fig:gapped}, but with a finite-width horizon, $\sigma_x = 1.9\xi_u$. Snapshots of the correlations have been taken at the same time instants described in Fig.~\ref{fig:gapless_time_dep_dx2}.  }
    \label{fig:gapped_time_dep_dx2_dx4}
\end{figure*}

We propagate each spin mode by numerically solving the Bogoliubov equations in Eq.~\eqref{Eq:GPE_App} according to a standard fourth-order Runge-Kutta scheme. To this end, we evaluate, at each time instant, the Bogoliubov operator on the numerical grid, according to the prescribed profile of $\kappa(x,t)$, $\Omega(x,t)$, as well as of the mean-field order parameter $\psi(x,t)$ (however, notice that the latter remains unchained since the chemical potential is kept constant). Given the solution for the modes, it is straightforward to construct any $\mathcal G_{\nu\nu'}(x,x')$ correlation function at the desired time. More details regarding the numerical solution are reported in Appendix \ref{appendix:AppBogotimedep}. 

Figs.~\ref{fig:gapless_time_dep_dx2} and \ref{fig:gapped_time_dep_dx2_dx4} display snapshots, at different times, of the correlation functions obtained with this technique, for the cases of zero and finite coupling in the downstream region, respectively. These results, obtained with a horizon of size $\sigma_x \approx 1.9\,\xi_s$, are in excellent qualitative agreement with the corresponding results in Figs.~\ref{fig:gapless} and \ref{fig:gapped}, which are obtained by using the scattering matrix approach in the assumption of infinitely steep horizon. 
The main quantitative difference between the two approaches resides in the smaller amplitude of the signal in the presence of a smooth horizon profile, which is, once again, imputed to the smaller Hawking temperature (see Fig.~\ref{fig:Hspectrum} and Appendix \ref{appendix:horizoneffect}).

\section{The critical point}
\label{sec:critical}
\begin{figure}[ht]
    \centering
    \includegraphics[width = 0.8\linewidth]{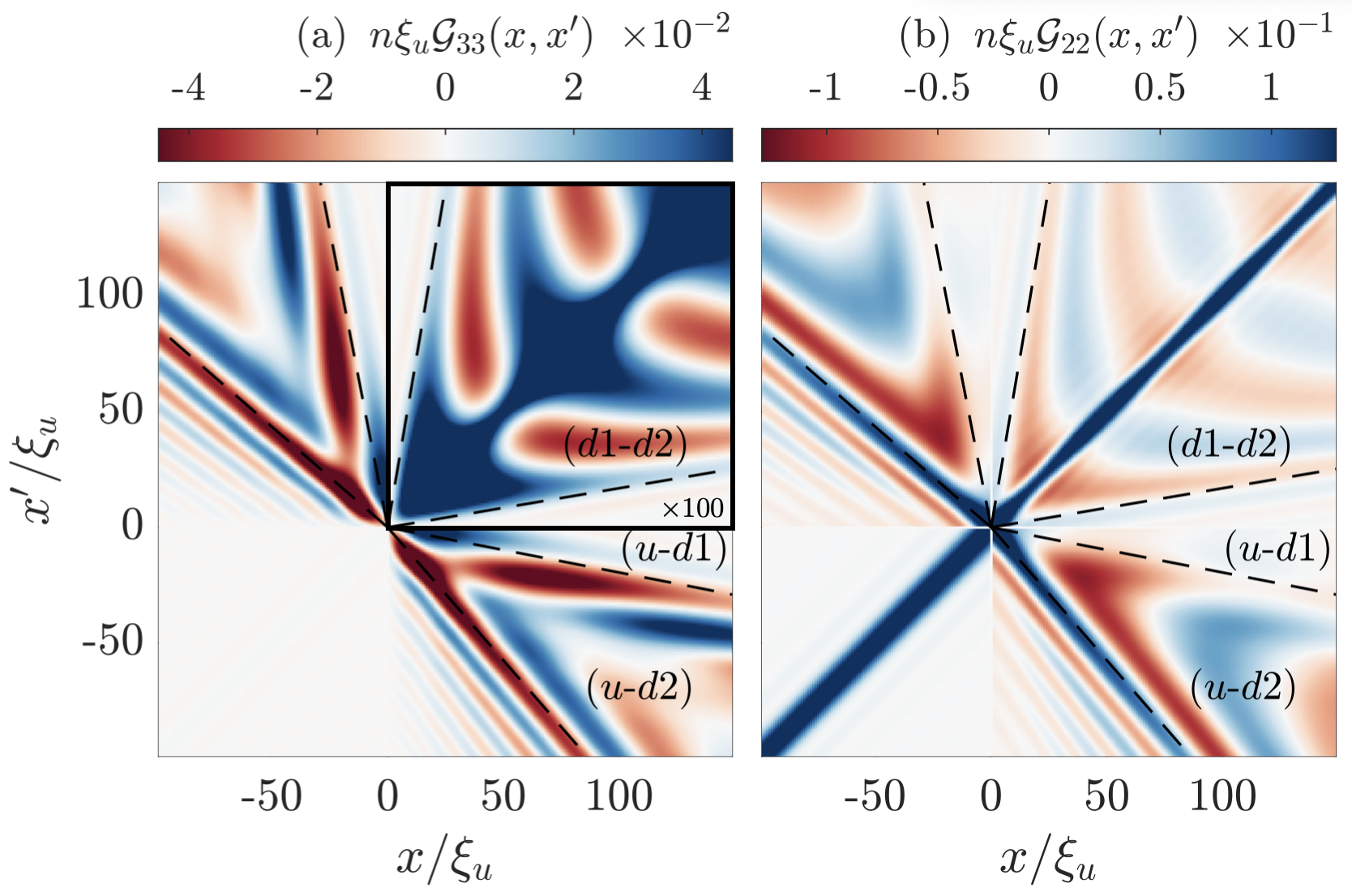}
    \caption{Density-density (a) and phase-phase (b) correlations with $\kappa_d/\kappa_u = -0.25$ and $\Omega_d = 1.004 |\kappa_d|n/\hbar$. The downstream region is therefore in the proximity of the para-to-ferromagnetic critical point. The signal for $\mathcal G_{33}$ in the $(x>0, x'>0)$ sector has been reduced of a factor 100, in order to show it on the same scale as the other contributions. }
    \label{fig:critical}
\end{figure}

Up to now, we have mostly considered parameters' regimes which are directly accessible to state-of-the-art experiments. In this Section, we focus instead on the conceptually interesting case defined by \eqref{eq:KGmassive2}: the system is close to the para-to-ferromagnetic phase transition point, thus relative density fluctuations are amplified. 
This regime is reached by setting, in the downstream region, $\kappa_d < 0$ and tuning the Rabi frequency $\Omega_d \gtrsim |\kappa_d|n$ so to get a small but finite gap $\Delta_d \ll |\kappa_d|n/\hbar$. Even though this case is not straightforwardly implemented in present set-ups it could be experimentally accessed in the next generation of experiments by exploiting spatially-dependent Feshbach resonances~\cite{chienSpatiallyVaryingInteractions2012}.  
In this peculiar configuration, zero-frequency modes have finite but small momentum $\pm k_0$, reducing the effect of undulations; at the same time, the large coupling in the downstream region regularizes the relative phase, making it possible to observe the standard Hawking mustache both in relative density and relative phase correlations. 
An example is given in Fig.~\ref{fig:critical}: for this choice of parameters, the signal is roughly one order of magnitude larger in $\mathcal G_{22}$ than in $\mathcal G_{33}$. 

\section{Conclusions and future perspectives}
\label{sec:concl}

We have studied analog Hawking emission of spin waves from a sonic horizon in a two-component Bose-Einstein condensate, both with and without a coherent coupling between the two components. 

We started from an analysis of the scattering properties of spin waves off the analog horizon, which carry information about the Hawking emission spectrum: while the presence of a coherent coupling in the downstream region leads to a modification of the greybody factor of the analog black hole, which acquires a quadratic frequency dependence in the limit $\omega\to 0$, we have verified that the thermal character of the emission of (massless) Hawking particles in the black hole exterior is not affected by the strength of the coherent coupling: this is remarkable considering that both the step configuration and the range of parameters we considered break the hydrodynamic approximation, and thus prevent a straightforward application of the gravitational analogy. 

We have then made use of the scattering matrix formalism to compute two-point correlation functions; in addition to density-density correlations, previously considered in the literature, we have computed phase-phase and density-phase correlation functions, with the aim of identifying the observable in which Hawking radiation manifests itself with largest intensity and, as such, is easiest to measure in a lab. We have first focused on the uncoupled two-component condensate, confirming the results of previous works \cite{larre2013hawking} and further extending them to prove the conceptual relevance of the phase degree of freedom in this configuration. 
Moreover, we considered the case of a coherently coupled two-component BEC and found that, in the limit of a large Rabi frequency in the downstream region ($\hbar\Omega_d \gg |\kappa_d|n$), phase correlations show an \textit{oscillating Hawking moustache}, characterized by the typical \textit{undulation} expected when massive particles are involved in the process; remarkably, the signal is orders of magnitude stronger than the one observed in density correlations for parameters within reach of currently available experiments \cite{cominotti2022observation}. 
All the semi-analytic predictions valid for the idealized case of a step configuration in an infinite system have been supported and validated by time-dependent numerical simulations based on the two-component Gross-Pitaevskii equation. This allowed us to investigate the effect of a smooth horizon profile and to verify the dynamical stability of the spin-sonic horizon configuration.

A natural extension of our work concerns the analysis of more realistic setups, such as the one in Fig.~\ref{fig:horizon_geometry}, involving a jump in the density profile and/or in the flow velocity, rather than in the spin interaction constant. 
Moreover, the robustness of the Hawking \textit{moustache} to a finite temperature needs to be investigated, especially in the presence of the coherent coupling. Along the same lines, a further development of our work could be the study of the effect of a small imbalance between the number of atoms in each component, that would lead to a weak coupling between density and spin modes \cite{cominotti2022observation}. 

Simple one-dimensional configurations similar to the one we considered throughout this paper are also suited, in principle, to analyse black-hole-related phenomena other than Hawking radiation. For instance, the addition of a finite coherent coupling in the upstream region might allow to observe a phenomenon known as the \textit{boomerang effect} \cite{jannes2011hawking}: modes with frequency smaller than the gap in the upstream region are emitted by the analog horizon but, instead of propagating to infinity, they bounce and are dragged back into the black hole interior. 

Another phenomenon that is expected to occur in black hole spacetimes is \textit{rotational superradiance} \cite{brito2020superradiance}: as already pointed out, the one-dimensional model considered in this work does not allow to observe superradiant amplification of upstream modes,  %because its analog geometry does not account for rotation, 
but superradiance-related phenomena could be analyzed in a planar configuration 
as soon as the fluid is set into rotation or
rotation is mimicked through a synthetic vector potential \cite{giacomelli2021understanding, giacomelli2021thesis}.
Analog geometries featuring both an horizon and an ergosurface would allow then to study the interplay between Hawking radiation, superradiant amplification, and the massive nature of modes \cite{brito2020superradiance}. 

Finally, it is worth noting that our results are not only relevant for atomic Bose-Einstein condensates, but also for polariton systems: in the latter case, two-component fluids are naturally generated by exploiting the polarization degree of freedom and a gap in the dispersion relation opens if the pump laser is detuned with respect to the polariton interaction energy \cite{jacquet2022analogue}.

\section*{Acknowledgments}
We thank Luca Giacomelli for fruitful discussions.

We acknowledge financial support by Provincia Autonoma di Trento (PAT) and from Q@TN, the joint lab between University of Trento, FBK-Fondazione Bruno Kessler, INFN-National Institute for Nuclear Physics and CNR-National Research Council. A. B. acknowledges funding from the European Union’s Horizon 2020 research and innovation Programme through the STAQS project of QuantERA II (Grant Agreements No. 101017733). I. C. acknowledges the support of the National Quantum Science and Technology Institute through the PNRR MUR Project under Grant PE0000023-NQSTI, co-funded by the European Union - NextGeneration EU.
L. F. was supported by a PhD fellowship of the Research Foundation - Flanders (FWO) under grant 11E8120N and a postdoctoral fellowship of the Belgian American Educational Foundation.
S. B. acknowledges funding from the Leverhulme Trust Grant No. ECF-2019-461, and from the University of Glasgow via the Lord Kelvin/Adam Smith (LKAS) Leadership Fellowship.
This work was supported in part through the NYU IT High Performance Computing resources, services, and staff expertise.

\bibliographystyle{crunsrt}

%This calls all references from the .bib
%\nocite{*}

%  This inserts the bib file
\bibliography{References}

\appendix
\include{appendix}

\end{document}

%% file: appendix.tex
\section{Bogoliubov theory}
\label{sec:App_A}
Let us consider a uniform $\mathbb{Z}_2$-symmetric mixture of total density $n$.
Bogoliubov spin eigenmodes of momentum $k$ and frequency $\omega(k)$ are typically expanded as the sum of a positive- and a negative-frequency contribution:
\begin{equation}
    \delta\psi(x,t) = U_k e^{ikx} e^{-i\omega t} + V_k^* e^{-ikx} e^{i\omega t}
    \label{eq:perturbation}
\end{equation}
where the coefficients $(U_k,V_k)$ are eigenstates of the Bogoliubov matrix:
\begin{equation}
    \mathcal L := \begin{bmatrix}
        \dfrac{\hbar^2k^2}{2m}+ \hbar\Omega  + \dfrac{\kappa n}{2} & \dfrac{\kappa n}{2} \\
        -\dfrac{\kappa n}{2} & -\left(\dfrac{\hbar^2k^2}{2m}+ \hbar\Omega  + \dfrac{\kappa n}{2} \right)
    \end{bmatrix}
    \label{eq:bogomatrix}
\end{equation}
namely $4U_kV_k = -\kappa n/\hbar\omega(k)$ and $(U_k+V_k)^2 = \mathcal S(k)$, where $\mathcal S(k)$ is the static structure factor:
\begin{equation}
    \mathcal S(k) = \sqrt{\frac{\hbar^2k^2/2m + \hbar\Omega}{\hbar^2k^2/2m + \hbar\Omega +\kappa n}}
    \label{eq:struct}
\end{equation}
One can also associate to each mode a conserved but not positive-definite norm, defined as:
\begin{equation}
    \avg{ \delta\psi |\sigma_3| \delta\psi } = |U_k|^2 - |V_k|^2 = \pm 1    \label{eq:Bogonorm}
\end{equation}
where $\sigma_3 = \text{diag}(1, -1)$ is the third Pauli matrix. The energy of the mode is given by the product of its norm and its frequency.

Let us also define, for later convenience:
\begin{align}
    R_k &= \sqrt{\mathcal{S}(k)}=U_k + V_k\\
    Q_k &= U_k - V_k=\pm R_k^{-1}
\end{align}

 Note that a mode with coefficients $(V_k^*, U_k^*)$, momentum $-k$ and frequency $-\omega$ corresponds to the same physical oscillation as \eqref{eq:perturbation} and carries the same energy content, due to the opposite sign of the Bogoliubov norm. In other words, the redundancy of the expansion \eqref{eq:perturbation} allows us to only consider modes with positive frequencies, provided we take into account modes with both signs of the norm. 

\section{The scattering solution}
\label{appendix:scattering}

Far from the sonic horizon ($|x|\to \infty$), the generic scattering solution at frequency $\omega$ is a eigenfunction of the spin Bogoliubov operator \eqref{eq:bogomatrix} of the form:
\begin{equation}
    \begin{bmatrix}
        \mathscr U (x) \\ \mathscr V(x)
    \end{bmatrix} = \sum_{r} \alpha_r  \begin{bmatrix}
        U_r^\text{in} \\ V_r^\text{in}
    \end{bmatrix} \frac{e^{ik_r^\text{in} x} }{ \sqrt{ |w_r^\text{in}|} }\Theta_r(x) + \sum_{r} \beta_{r}   \begin{bmatrix}
        U_{r}^\text{out} \\ V_{r}^\text{out}
    \end{bmatrix} \frac{e^{ik_{r}^\text{out} x}}{ \sqrt{|w_{r}^\text{out}|} } \Theta_{r}(x)
\label{eq:Bogosolution}
\end{equation}
where the sum runs over the three branches and includes all the propagating modes available at frequency $\omega$, separated into ingoing and outgoing ones.  In particular, $r \in \{u, d1, d2\}$ if $\omega < \omega_*$ and $r \in \{u, d1\}$ if $\omega > \omega_*$. 

In the vicinity of the horizon ($x \sim 0$), one must include exponentially suppressed modes in the expansion by adding to Eq.\eqref{eq:Bogosolution} a term of the form:
\begin{equation}
    + \sum_r  \gamma_r  \begin{bmatrix}
        U_r^\text{ev} \\ V_r^\text{ev}
    \end{bmatrix} e^{ik_r^\text{ev} x} \Theta_r(x) 
\end{equation}
where $r \in \{u\}$ if $\omega < \omega_*$ and $r \in \{u, d\}$ if $\omega > \omega_*$.

The Heaviside Theta functions $\Theta_r(x)$ are necessary to ensure that the mode $u$ (the modes $d1,d2$) contributes to the solution only in the upstream (downstream) region $x<0$ ($x>0$):
\begin{equation}
    \Theta_u(-x) = \Theta_{d1}(x) = \Theta_{d2}(x) = \begin{cases}
        0 & \text{if } x < 0 \\
        1 & \text{if } x > 0
    \end{cases}
\end{equation}
The normalization involving the group velocity guarantees that the orthonormality condition is fulfilled in frequency space. 
The particle and antiparticle components ($U_k, V_k$) are solutions of the Bogoliubov problem in the comoving frame; as such
\begin{equation}
    |U_r^\text{in}|^2 - |V_r^\text{in}|^2 = |U_r^\text{out}|^2 - |V_r^\text{out}|^2 = \sigma_r
\end{equation}
where $\sigma_u = \sigma_{d1} = - \sigma_{d2}=+1$ is the Bogoliubov norm.

Thanks to the relation between in-going and out-going modes coefficients $\alpha_r, \beta_r$ expressed by the scattering matrix, the generic scattering solutions \eqref{eq:Bogosolution} can be written as a linear combination of scattering modes. We call \textit{scattering mode} the solution of the Bogoliubov problem initiated by a single in-going quasi-particle mode:
\begin{equation}
    \begin{split}
        \Bigg[ \begin{matrix}
        \mathcal U_r (x) \\ \mathcal V_r (x)
    \end{matrix}\Bigg] &= \begin{bmatrix}
        U_r^\text{in} \\ V_r^\text{in}
    \end{bmatrix} \frac{e^{ik_r^\text{in} x} }{  |w_r^\text{in}|^{1/2} }\Theta_r(x) +\sum_{r'} \mathcal M_{r'r}   \begin{bmatrix}
        U_{r'}^\text{out} \\ V_{r'}^\text{out}
    \end{bmatrix} \frac{e^{ik_{r'}^\text{out} x}}{ |w_{r'}^\text{out}|^{1/2} } \Theta_{r'}(x) 
    \end{split}
    \label{eq:scattering_in}
\end{equation}

The computation of the scattering matrix coefficients typically requires to fully solve the Bogoliubov problem, taking into account the exact profile of the sonic horizon. In analogy to the single-component case \cite{recati2009bogoliubov}, a semi-analytical solution can be easily determined for the step-like configuration in Fig.~\ref{fig:sketch}. 
Indeed, physical profiles for the Bogoliubov amplitudes \eqref{eq:Bogosolution} are found by matching the solutions in the subsonic and supersonic regions at the location of the horizon. 
In practice, the amplitudes $\alpha_r$ of the in-going modes are fixed as initial conditions for the scattering problem, while those of the out-going modes $\beta_r$ are computed by matching the $x=0$ values of $\mathcal U, \mathcal V$ and of their first derivatives. This amounts to solving a linear system of four equations for each in-going mode an for each frequency $\omega$. Since the matching is done at the position of the horizon, one must include exponentially suppressed modes in the calculations; from a mathematical point of view, this is necessary to get unambiguous results, since the number of unknown coefficients to be determined ($\beta_r$ and $\gamma_r$) needs to be equal to the number of imposed constraints.

\section{Quantization of the field}
\label{appendix:quantization}
The standard quantization procedure \cite{recati2009bogoliubov} consists in replacing complex amplitudes with creation/annihilation operators. The field perturbation can thus be written in terms of scattering modes \eqref{eq:scattering_in} as:
\begin{equation}
    \hat{\delta\Psi}(x,t) = \int_0^\infty \frac{\text{d}\omega}{\sqrt{2\pi}} \sum_r \Big[ \mathcal U_r(x) e^{-i\omega t} \hat a_r + \mathcal V^{*}_r(x) e^{i\omega t} \hat a_r^\dagger \Big]
    \label{eq:fieldpert}
\end{equation}
where the sum runs over all the modes available at a given frequency, that is $r\in \{u, d1,d2\}$ if $\omega < \omega_*$ and $r\in \{u, d1\}$ if $\omega > \omega_*$.
Notice that Eq.\eqref{eq:fieldpert} is not exact in the vicinity of the analog horizon ($x\ne 0$), since evanescent modes have not been included. 
Also, we have associated annihilation (creation) operators to the $U$ ($V$) component for all the modes; with this convention, in order to guarantee the proper commutation relations for the field, we have to treat negative norm modes as anti-particles, for which the roles of creation and annihilation operators are exchanged. This reflects on the commutation relations:
\begin{align}
    [\hat a_r(\omega), \hat a_{r'}^\dagger(\omega')] &= \sigma_r \delta_{rr'}\delta(\omega-\omega')
\end{align}
as well as on the expectation values. 
If the symbol $\avg{\cdot}$ indicates the average value over the vacuum of in-going particles, at zero physical temperature $T=0$ we have:
\begin{align}
    \avg{\hat a_r^\dagger(\omega) \hat a_{r'}(\omega')} &=  \Big( \frac{1-\sigma_r}{2}\Big) \,\delta_{rr'}\delta(\omega-\omega') 
    \label{eq:commutators_a}\\
    \avg{\hat a_{r'}(\omega') \hat a_r^\dagger(\omega) } &=  \Big( \frac{1+\sigma_r}{2}\Big)\delta_{rr'}\delta(\omega-\omega') 
    \label{eq:commutators_b}
\end{align} 

In principle, in order to expand $\delta\Psi$ over a complete basis set, one would need to explicitly include two unconventional eigenstates of the Bogoliubov matrix $\mathcal L$ with eigenvalue $\omega = 0$ \cite{isoard2020departing}. In a single component BEC, these zero-frequency modes are 
due to the Gauge symmetry associated with the global phase of the system, or, equivalently, to the conservation of the total number of particles. In the case of the spin channel of a symmetric binary mixture, they exist only in the absence of a coherent coupling, when the relative number of particles is conserved and the same freedom applies to the relative phase. Within our framework, involving a uniform density profile, the Bogoliubov coefficients of these two zero-frequency modes would also be position-independent. As a consequence, without coherent coupling, they would contribute, at most, as a uniform shift to the observables we will define in the following; if $\Omega \ne 0$ in the downstream region, the continuity of the wavefunction fixes the relative phase in the upstream region as well, effectively eliminating the Goldstone mode. For these reasons, we avoid including the zero-frequency modes in the field expansion \eqref{eq:fieldpert} and in the following calculations.
This affects, however, the field commutator, which no longer coincides with a Dirac delta function:
\begin{equation}
\begin{split}
    [\hat{\delta\Psi}(x), \hat{\delta\Psi}^\dagger(x')] 
    &= \int_0^\infty \frac{\text{d}\omega}{2\pi} \sum_{r}  \frac{e^{ik_r^\text{in}(x-x')}}{|w_r^\text{in}|}  \Theta_r(x)\Theta_{r}(x') + \int_0^\infty \frac{\text{d}\omega}{2\pi} \sum_{r}  \frac{e^{ik_r^\text{out}(x-x')}}{|w_r^\text{out}|}  \Theta_r(x)\Theta_{r}(x')
\end{split}
    \label{eq:fieldcomm}
\end{equation}

Starting from the expansion of the field perturbation \eqref{eq:fieldpert} in terms of scattering modes, we can derive the second and third component of the spin vector \eqref{eq:spinvector}:
\begin{align}
\hat S_2(x,t) &= i \sqrt n \big( \hat{\delta\Psi}-\hat{\delta\Psi}^\dagger \big) = \sqrt n \int_0^\infty \frac{\text{d}\omega}{\sqrt{2\pi}} \sum_{r} \left[ i \mathcal Q_r(x)e^{-i\omega t} \hat a_r  + \text{h.c.} \right] \label{eq:spinvector2_expansion}  \\
\hat S_3(x,t) &= \sqrt n \big( \hat{\delta\Psi}+\hat{\delta\Psi}^\dagger \big) = \sqrt n \int_0^\infty \frac{\text{d}\omega}{\sqrt{2\pi}} \sum_{r} \left[ \mathcal R_r(x) e^{-i\omega t} \hat a_r  + \text{h.c.} \right] \label{eq:spinvector3_expansion} 
\end{align}
where, in order to simplify the notation, we have defined the sum and difference of the Bogoliubov components describing the scattering modes as 
\begin{align*}
    \begin{split}
        \mathcal R_r (x) &= \mathcal U_r (x)+\mathcal V_r (x) = \frac{R_r^\text{in} e^{ik_r^\text{in} x} }{  |w_r^\text{in}|^{1/2} }\Theta_r(x) + \sum_{r'} \mathcal M_{r'r}   \frac{R_{r'}^\text{out} e^{ik_{r'}^\text{out} x}}{ |w_{r'}^\text{out}|^{1/2} } \Theta_{r'}(x) 
    \end{split} \\
    \begin{split}
        \mathcal Q_r (x) &= \mathcal U_r (x)-\mathcal V_r (x) = \frac{Q_r^\text{in} e^{ik_r^\text{in} x} }{  |w_r^\text{in}|^{1/2} }\Theta_r(x) + \sum_{r'} \mathcal M_{r'r}   \frac{Q_{r'}^\text{out} e^{ik_{r'}^\text{out} x}}{ |w_{r'}^\text{out}|^{1/2} } \Theta_{r'}(x) 
    \end{split}
\end{align*}

\section{Two-point correlations}
\label{appenxix:corr}

Equal-time two-point correlations [defined in the main text in Eqs.\eqref{eq:G33definition}, \eqref{eq:G22definition} and \eqref{eq:G23definition}] can be computed at zero temperature by exploiting the expansion of the spin vector component in terms of scattering modes \eqref{eq:spinvector2_expansion}, \eqref{eq:spinvector3_expansion}, the expectation values \eqref{eq:commutators_a}, \eqref{eq:commutators_b} and the properties of the scattering matrix. A direct calculation gives:
\begin{align}
    \mathcal G_{33}(x,x') &= \frac{\avg{\hat S_3(x)\hat S_3(x')}}{n^2}- \frac 1 n \big[\hat{\delta\Psi}(x), \hat{\delta\Psi}^\dagger(x')\big]\notag \\
    &= \Re \int_0^\infty \frac{\text{d}\omega}{2\pi} \sum_r \left[ \frac{ |R^\text{in}_{r} |^2 -1}{n |w_r^I|} e^{ik_r^\text{in}(x-x')} + \frac{ |R^\text{out}_{r} |^2 -1}{n |w_r^I|} e^{ik_r^\text{out}(x-x')} \right] \Theta_r(x)\Theta_{r}(x') \label{eq:corr33diag} \\
    &\qquad + \Re \int_0^{\omega_*} \frac{\text{d}\omega}{2\pi} \sum_{rr'} \left[ \frac{ R_{r}^\text{out} R_{r'}^{\text{out}} e^{i(k_r^\text{out}x-k_{r'}^\text{out}x')} } {n |w_r^\text{out}w_{r'}^\text{out}|^{1/2}} \left( \mathcal M_{rd2}\mathcal M_{r'd2}^* + \frac{\sigma_r-1}{2} \delta_{rr'} \right) \right] \Theta_r(x)\Theta_{r'}(x') \label{eq:corr33hawking} \\
    &\qquad\qquad \notag + (x \leftrightarrow x') 
\end{align}
\begin{align}
    \mathcal G_{22}(x,x') &= \frac{\avg{\hat S_2(x)\hat S_2(x')}}{n^2} - \frac 1 n \big[\hat{\delta\Psi}(x), \hat{\delta\Psi}^\dagger(x')\big]\notag \\
    &= \Re \int_0^\infty \frac{\text{d}\omega}{2\pi} \sum_r \left[ \frac{ |Q^\text{in}_{r} |^2 -1}{n |w_r^I|} e^{ik_r^\text{in}(x-x')} + \frac{ |Q^\text{out}_{r} |^2 -1}{n |w_r^I|} e^{ik_r^\text{out}(x-x')} \right] \Theta_r(x)\Theta_{r}(x') 
    \label{eq:corr22diag} \\
    &\qquad + \Re \int_0^{\omega_*} \frac{\text{d}\omega}{2\pi} \sum_{rr'} \left[ \frac{ Q_{r}^\text{out} Q_{r'}^{\text{out}} e^{i(k_r^\text{out}x-k_{r'}^\text{out}x')} } {n |w_r^\text{out}w_{r'}^\text{out}|^{1/2}} \left( \mathcal M_{rd2}\mathcal M_{r'd2}^* + \frac{\sigma_r-1}{2} \delta_{rr'} \right) \right] \Theta_r(x)\Theta_{r'}(x') \label{eq:corr22hawking} \\
    &\qquad\qquad \notag + (x \leftrightarrow x') \\
    \mathcal G_{23}(x,x') &= \frac 1 2 \left[ \frac{\avg{\hat S_2(x)\hat S_3(x')}+\avg{\hat S_3(x)\hat S_2(x')}}{n^2} \right]\notag \\
    &= \Im  \int_0^{\omega_*} \frac{\text{d}\omega}{2\pi}  \sum_{r\ne r'} \left[  \frac{  R_{r}^\text{out} Q_{r'}^{\text{out}}e^{i(k_r^\text{out}x-k_{r'}^\text{out}x')} } {n |w_r^\text{out}w_{r'}^\text{out}|^{1/2}}  \mathcal M_{rd2}\mathcal M_{r'd2}^* \right] \Theta_r(x)\Theta_{r'}(x') \label{eq:corr23hawking}\\
    &\qquad\qquad  + (x\leftrightarrow x') \notag 
\end{align}
The semi-analytical calculation of the correlation signals shown in the main text in Figs.\ref{fig:gapless} and \ref{fig:gapped} is performed with formulas \eqref{eq:corr33hawking}, \eqref{eq:corr22hawking} and \eqref{eq:corr23hawking}. 

\begin{figure}[h]
    \centering
    \includegraphics[width = 0.7\linewidth]{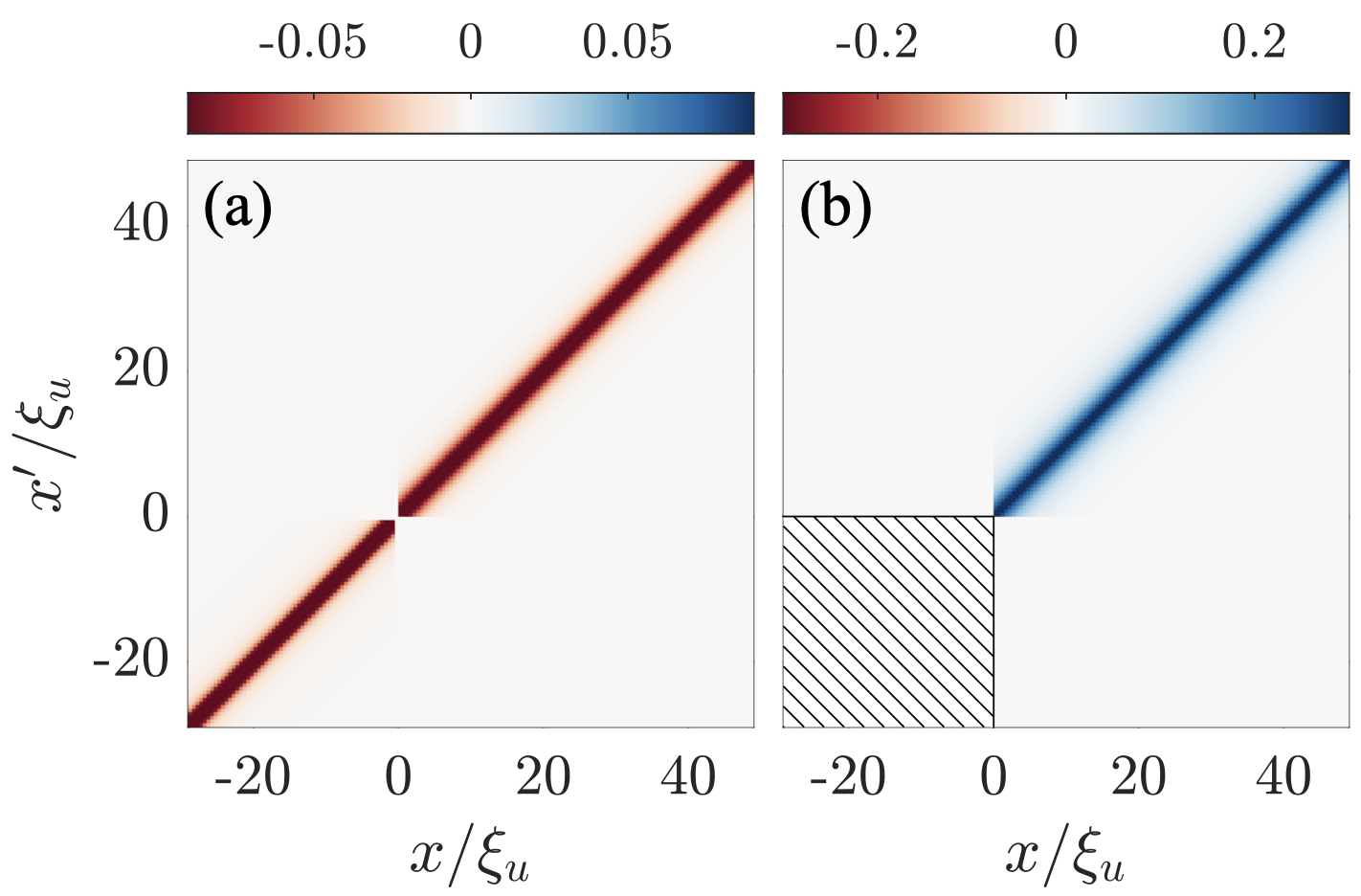}
    %\tikz \draw (0,0) rectangle (\linewidth, 0.7\linewidth);
    \caption{Trivial contribution to the relative density (a) and relative phase (b) correlation signal due to spin interactions, computed using Eqs. \eqref{eq:corr33diag} and \eqref{eq:corr22diag}, respectively. Parameters are identical to those in Fig.~\ref{fig:S_omega_ideal} ($\Omega_d\ne 0$).  Notice that our calculation predicts relative phase correlations to diverge in the upstream region (hatched area of the plot), due to the gapless nature of excitations. However, if the coherent coupling is present in the downstream region, it intrinsically fixes a reference value (0) for the relative phase. This has the same regularization effect of measuring the phase with respect to the analog horizon.   }
    \label{fig:corrdiag}
\end{figure}

Consistently with \cite{larre2012quantum}, 
relative density correlations consist in a sum of various contributions: the first trivial term \eqref{eq:corr33diag} is present even in the absence of an analog horizon and  represents an antibunching (bunching) term resulting from repulsive (attractive) spin interactions.  When expanded in $k$-space rather than frequency space, it coincides with the Fourier transform of the structure factor reduced by 1, computed at $x-x'$: as a consequence it diverges at the critical point for the para-to-ferromagnetic phase transition, where $\mathcal S(k\to 0) \propto 1/k$, while it vanishes in the absence of spin interactions, $\kappa = 0$, since $\mathcal S(k) = 1$. A typical example is given in Fig.~\ref{fig:corrdiag}(a): for $\kappa > 0$, the signal consists a negative correlation peaked along the main diagonal $x=x'$; the difference in the signal found in the two sectors $x,x'>0$ and $x,x'<0$ is due to the different value of spin interaction energy $\kappa$ in the upstream and downstream regions.    

The additional correlation term in Eq.\eqref{eq:corr33hawking} contributes only if the $d2$ mode is available, that is, only in the presence of a sonic horizon. For this reason, it can be directly associated to the analog of the Hawking effect in a BEC system. We can distinguish between self-correlation ($r=r'$) and cross-correlation $(r\ne r')$ terms: while the former give, once again, a signal along the main diagonal $x=x'$, the latter can contribute away from it, producing a signal which is clearly distinguishable from the usual bunching/antibunching term. In particular cross-correlations are non-vanishing in the quadrants $xx'< 0$: physically such correlation signal is interpreted as due to pairs of quasi-particles which are spontanously emitted on the two sides of the sonic horizon, in close analogy with what is expected in gravitational context. 

Analogously to density correlations, also phase correlations can be written as the sum of various contributions, including a trivial term \eqref{eq:corr22diag} due to spin interactions: this consists in a positive (negative) contribution for $\kappa > 0$ ($\kappa < 0$), localized around the main diagonal $x=x'$.
When expanded in $k$-space, it coincides with the Fourier transform of the inverse of the structure factor reduced by 1: therefore it diverges in the absence of a coherent coupling, while it vanishes in the absence of spin interactions. A typical example is given in Fig.~\ref{fig:corrdiag}(b). 

The additional contributions \eqref{eq:corr22hawking}, which are only found if an analog horizon is present, can again be distinguished between self-correlations ($r=r'$) and cross correlations ($r\ne r'$). Once again, only the latter are expected to produce a signal that is clearly distinguishable from the trivial contribution along the main diagonal. In particular, the $(ud)$ contributions in the sectors with $xx'<0$ are associated to correlations between the Hawking particles emitted in the black-hole exterior and their negative energy partners. 

The result for mixed density-phase correlations \eqref{eq:corr23hawking} is qualitatively different: the real part appearing in Eq.\,\eqref{eq:corr22hawking}, \eqref{eq:corr33hawking} is replaced by an imaginary part due to the imaginary unit in the expansion of $S_2$, see Eq.\,\eqref{eq:spinvector2_expansion}. 
As a consequence, all self-correlation terms vanish and $\mathcal G_{23}(x,x')$ is exclusively given by the cross-correlations. In other words, this mixed correlation signal is entirely due to the Hawking process and would be absent without an analog horizon. 

\subsection{Gapless case}
All the main features appearing in the two-point correlation functions can be explained through approximated analytical calculations of the above integrals in the low-frequency regime $\omega\to 0$. The momenta of the out-going modes can be considered linear with respect to $\omega$:
\begin{equation}
    k_r^\text{out}(\omega \to 0) \simeq \omega/w_r^\text{out}(0)
    \label{eq:linearwavevector}
\end{equation}
As a consequence, the phase factor
\begin{align*}
    \Phi_{rr'}(x,x', \omega) &\equiv k_r^\text{out}(\omega)x - k_{r'}^\text{out}(\omega)x' \simeq \omega \bigg(\frac{x}{w_r^\text{out}(0)}- \frac{x'}{w_{r'}^\text{out}(0)}\bigg) \equiv \omega \Pi_{rr'}(x,x')
\end{align*}
is also approximatively linear in $\omega$, as well as the structure factor for spin modes $\mathcal S(k_r^\text{out})$; therefore we have:
\begin{align}
    R_{r}^\text{out}(\omega \to 0) &\propto \sqrt{\omega} \\
    Q_{r}^\text{out}(\omega \to 0) &\propto \sigma_r/\sqrt{\omega}
\end{align}
Moreover, in the low-frequency regime, the solution of the scattering problem gives the following results for the scattering matrix elements:
\begin{align}
    \Re \big[\mathcal M_{rd2}(\omega \to 0)\mathcal M_{r'd2}^*(\omega \to 0)\big] &\propto 1/\omega \\
    \Im \big[\mathcal M_{rd2}(\omega \to 0)\mathcal M_{r'd2}^*(\omega \to 0)\big] &\simeq \text{const}
\end{align}
so that we can safely neglect their imaginary part in the following calculation.

Putting all together, we find that the main contribution to the density correlation signal $\mathcal G_{33}(x,x')$ is a sum of oscillating functions, one for each pair of modes:
\begin{align*}
    \mathcal G_{33}(x,x') &\propto \int_0^{\omega_*} \cos\left[ \omega \Pi_{rr'}(x,x') \right]\text{d}\omega \propto  \text{sinc}\big[\omega_* \Pi_{rr'}(x,x')\big]
    \label{eq:G33_gapless}
\end{align*}
where sinc$(z) \equiv \sin(z)/z$. 
Each of these terms is peaked at the locus of points fulfilling $\Pi_{rr'}(x,x')=0$, that is, on the straight lines defined by Eq. \eqref{eq:corrlines_gapless}.

Following the same procedure, we can also compute mixed density-phase correlations $\mathcal G_{23}(x,x')$ and phase correlations $\mathcal G_{22}(x,x')$. The main contribution to the former is given by terms of the form:
\begin{align*}
    \mathcal G_{23}(x,x') &\propto \int_0^{\omega_*} \frac{\sin[\omega \Pi_{rr'}(x,x')]}{\omega}\text{d}\omega \propto \text{Si} \big[ \omega_*\Pi_{rr'}(x,x') \big]
\end{align*}
where Si$(z) = \int_0^z \text{sinc}(z') \text{d} z'$ is the sine integral function: its asymptotic values at $\pm \infty$ are $\pm \pi/2$, respectively, and it crosses zero when $z=0$. The full correlation, shown in Fig.~\ref{fig:gapless}(b), is therefore a collection of patches in which $\mathcal G_{23}$ has different, almost constant values, separated by smooth jumps whose location is determined, once again, by the solution of $\Pi_{rr'}(x,x')=0$. 

Lastly, phase-phase correlations $\mathcal G_{22}$ show an infrared divergence: indeed the $1/\omega$ behaviour of the scattering matrix coefficients is not compensated and even worsened by the $Q_{r}^\text{out}$ coefficients, which provide a factor which is also inversely proportional to the frequency. Given some infrared cut-off $\epsilon$, the most relevant contribution is:
\begin{equation}
    \mathcal G_{22}(x,x') \propto \int_\epsilon^{\omega_*} \frac{\cos[\omega \Pi_{rr'}(x,x')]}{\omega^2}\text{d}\omega 
    \label{eq:gapless_corrph}
\end{equation}
and diverges as $1/\epsilon$. The trivial phase correlations in Eq.\eqref{eq:corr22diag} also diverge in the absence of coherent coupling, but in a slower way, as $\log(1/\epsilon)$. As already discussed in the main text, the infrared divergence of phase correlations can be cured by measuring $\hat S_2$ with respect to some reference point (the sonic horizon $x=0$ is the most natural choice). This amounts to substituting, in formulas \eqref{eq:corr22diag}, \eqref{eq:corr22hawking}:
\begin{equation}
    e^{i(k_r^\text{out}x-k_{r'}^\text{out}x')} \longrightarrow \big(e^{ik_r^\text{out}x}-1\big)\big(e^{ik_{r'}^\text{out}x'}-1\big) 
    \label{eq:regularizephase}
\end{equation}
In the low-frequency limit, the complex exponentials \eqref{eq:regularizephase} can be expanded at first order in $\omega$ and the phase factor $e^{i\omega\Pi(x,x')}$ is replaced by $k_r^\text{out}k_{r'}^\text{out} xx'\propto \omega^2 xx'$: the $\omega^2$ dependence cancels the denominator in Eq.\eqref{eq:gapless_corrph}, leaving a finite correlation signal $\mathcal G_{22}(x,x') \propto xx'$. 

Notice that, in principle, Eq. \eqref{eq:spinvector2_expansion} is only valid away from the sonic horizon because it does not include evanescent modes. However, these spatially decay over distances of a few healing lengths, and long range coherence in BEC system guarantees that the error we make by estimating the phase at $x=0$ with Eq.\eqref{eq:spinvector2_expansion} is negligible.

\subsection{(Non-relativistic) gapped case} 
Let us now repeat the same low-frequency approximated calculation in the presence of a large coherent coupling in the downstream region only. Since the gap is only present in the downstream region, we have to treat separately the $u$ mode and the $d = \{d1,d2\}$ modes. A Taylor expansion of the out-going momenta (up to linear terms in $\omega$) leads to:
\begin{align*}
    k_u^\text{out}(\omega\to 0) &\simeq \omega/w_u^\text{out}(0)\\
    k_d^\text{out}(\omega\to 0) &\simeq \pm k_0 + \omega/w_0 
\end{align*}    
where $k_0, w_0$ are the momentum and group velocity of zero-frequency modes in the downstream region. The structure factors are approximately given by:
\begin{align*}
    \mathcal S(k_u^\text{out}) &\propto \omega \\
    \mathcal S(k_d^\text{out}) &\simeq \text{const} 
\end{align*}
This translates on the phase factors as follows:
\begin{align*}
    \Phi_{uu}(x,x', \omega) &\simeq \omega \,\frac{(x-x')}{w_u^\text{out}(0)} \\
    \Phi_{ud}(x,x', \omega\to 0) &= \omega \Pi_{ud}(x,x') \pm k_0 x' \\
    \Phi_{dd'}(x,x', \omega\to 0) & = \pm k_0 ( x \pm x') + \omega \frac{(x-x')}{w_0} \label{eq:Phidd}
\end{align*}
In general, the phase factor contains a frequency-independent contribution, which will produce an oscillating correlation signal with momentum $k_0$, as well as a term proportional to $\omega$, that determines the locus of points in the $(x,x')$ plane in which the signal is peaked: self-correlations contribute only close to the main diagonal.  The same holds for $(d1\text{-}d2)$ cross-correlations, due to the identical group velocity for the two modes. On the contrary, $(u\text{-}d1)$ and $(u\text{-}d2)$ correlations are expected to be superimposed on a line defined by:
\begin{equation}
    \Pi_{ud}(x,x') = \frac{x}{w_u^\text{out}(0)} - \frac{x'}{w_0} = 0
    \label{eq:udcorrlines}
\end{equation}

The solution of the scattering problem gives the following results for the scattering matrix coefficients: those involving $d$ modes tend to a constant value, while $\mathcal M_{ud2}$ tends to zero in the zero-frequency limit: 
\begin{align}
    |\mathcal M_{ud2}(\omega\to 0)|^2 &\propto \omega \notag \\
    \mathcal M_{ud2}(\omega\to 0)\mathcal M^*_{dd2}(\omega\to 0) &\propto \sqrt{\omega} \label{eq:Mzerofreq} \\
    \mathcal M_{dd2}(\omega\to 0)\mathcal M^*_{d'd2}(\omega\to 0) &\simeq \text{const} \notag
\end{align}
These asymptotic results allow us to estimate the features appearing in two-point correlations. Let us focus on phase correlations $\mathcal G_{22}(x,x')$, and, more specifically, on the $(ud)$ and $(dd)$ sectors, where the signal is espected to be significantly different with respect to the trivial contribution in Fig.~\ref{fig:corrdiag}. 
Cross correlations between $u$ and $d$ modes give, in the quadrants $xx'<0$:
\begin{equation*}
    \mathcal G_{22}(x,x') \propto \cos(k_0 x) \text{sinc}\big[ \omega_*\Pi_{ud}(x,x') \big] 
\end{equation*}
while in the $(dd)$ quadrant ($x>0,x'>0$) we find a checkerboard pattern peaked on the main diagional, arising from the interference between $(d1$-$d2)$ cross-correlations and self-correlations of downstream modes: 
\begin{equation*}
    \mathcal G_{22}(x,x') \propto \cos(k_0 x)\cos(k_0x') \text{sinc}\left[ \omega_* \frac{(x-x')}{w_0} \right]
\end{equation*}
This checkerboard pattern has the same origin of the one found for acoustic white-hole configurations \cite{mayoral2011acoustic}, namely the presence of zero-frequency out-going modes with non-zero momentum. However, thanks to the regularization of the scattering amplitudes, it is not infrared divergent. 
These simple results are in good agreement with the correlation signal shown in Fig.~\ref{fig:gapped}. 

%\bigskip
%\anna{The semi-analytical calculations we have shown so far assume the state of the system is stationary, while realistic set-ups only allow to measure correlations at finite times $t$ after the creation of the analog horizon. While an accurate analysis of the two-point correlations at short and intermediate times requires time-dependent numerical simulations, a rough estimate of how these grow with time can be achieved by artificially setting an infrared frequency cutoff $\epsilon = 1/t$.  }

%\anna{In Fig.~\ref{fig:max_gapped} we report the maximum value that the same phase correlations shown in Fig.~\ref{fig:gapped} take as a function of the cutoff frequency: it is clear that the signal saturates for $\hbar\epsilon/mc_u^2 < 10^{-3}$, that is, for times longer than $10^3 mc_u^2/\hbar$. At shorter time scales the correlation signal might then appear to be growing in time and potentially have a different structure than the one discussed and shown in Fig.~\ref{fig:gapped}, that we computed with $\epsilon = 10^{-5} mc_u^2/\hbar$. }

%\begin{figure}[H]
%    \centering
%    \includegraphics[width = 0.65\linewidth]{Figures/max_gapped.pdf}
%    \caption{\anna{Maximum value acquired by phase correlations as a function of an artificially imposed frequency infrared cutoff. Same parameters as in Fig.~\ref{fig:gapped}. } }
%    \label{fig:max_gapped}
%\end{figure}

\section{Extracting the scattering properties of the spin-sonic horizon from time-dependent GP simulations}
\label{appendix:GPsim}

As the spectrum \eqref{eq:Bogodisp} of Bogoliubov excitations underlying the Hawking effect equally arises for small amplitude oscillations on a steady state solution of the GP equations \eqref{Eq:GPE} \cite{pitaevskii2016bose}, classical scattering properties of the spin-sonic horizon reported in Figs. \ref{fig:Bragg_example}, \ref{fig:S_omega_ideal} and \ref{fig:Hspectrum} are inferred by numerically integrating the coupled GP equations \eqref{Eq:GPE} through a split-operator method \cite{wangTimesplittingSpectralMethod2007, sartorialbertoDynamicalPropertiesBoseBose2016}.
To mimic the idealized configuration outlined in Sec. \ref{sec:spinsonic}, the system is initialized in the uniformly flowing state \eqref{eq:hawking_stationarystate} in a periodic geometry, divided by a spin-sonic horizon resulting from a modulation of the spin interaction strength\footnote{By contrast, the non-uniform density profile of the realistic implementation shown in Fig.~\ref{fig:horizon_geometry} introduces additional complexity by inducing a modulation of the density interaction energy $(g + g_{\uparrow \downarrow})n$.}.
As noted in the main text, the idealized configuration of a sharp modulation of the interaction energy \eqref{eq:mu_modulation} is generalized to a smooth step profile of width $\sigma_x$,
\begin{equation}
    \kappa(x) = \frac{\kappa_u + \kappa_d}{2} + \qty(\frac{\kappa_d - \kappa_u}{2})\tanh(\frac{x}{\sigma_x}). \label{eq:smooth_horizon_profile}
\end{equation}
Besides allowing for a tuning of the Hawking temperature \eqref{eq:analogTH} as discussed below, a smooth horizon profile was found to be required to ensure numerical stability of the time-dependent GP equations in the case of a non-uniform coherent coupling $\Omega$ across the horizon. Such instability is briefly discussed in Sec.\ref{appendix:horizoneffect} and will be the subject of a future work.

A scattering process is initiated through the localized excitation of a spin mode at time $t_0$ and position $x_0$ by a pulse of the form
\begin{equation}
V(x,t) = \pm A \exp(-\frac{(t-t_0)^2}{2\Delta_t^2}-\frac{(x-x_0)^2}{2\Delta_x^2})\cos(k x - \omega t),
\label{eq:bragg_pulse}
\end{equation}
where the wavelength $1/k$ and frequency $\omega$ are tuned to be on resonance with an in-going branch of the upstream \eqref{eq:disp_up_labframe} or downstream \eqref{eq:disp_down_labframe} spin dispersion relation in the lab frame, and the sign $\pm$ refers to the two components $\psi_{\uparrow, \downarrow}$, respectively.
The amplitude $A$ is tuned to ensure all spin excitations remain in the linear response regime, i.e.,
\begin{equation}
    \delta \psi = \frac{\psi_\uparrow - \psi_\downarrow}{\psi_\uparrow + \psi_\downarrow} \leq 10^{-3}. \label{eq:GPE_spin_response}
\end{equation}
 
Wave packets of sufficient size to infer spectral properties of the Hawking process in the long-wavelength limit are generated by pulses of spatiotemporal profile $\Delta_t, \Delta_x \sim 10(\kappa_u n)^{-1}, 10^3\xi_u$, while the size of the system itself is restricted by computational complexity\footnote{The limited spatial extent of the system manifests mostly in the scattering of $u$-ingoing modes, the momenta of which approach $k\to 0$ in the limit $\omega \to 0$, resulting in non-negligible noise visible in Fig.~\ref{fig:Hspectrum}(b).} to the order $10^5\xi_u$ on a discretized grid of resolution $\delta x=\xi_u/4$.
Following a scattering process as depicted in Fig.~\ref{fig:Bragg_example}, the spin response \eqref{eq:GPE_spin_response} evaluated in the comoving frame consists of several spatially separated wave packets, each of which is a product of a Gaussian envelope of width $|w_{r'}^{out}(\omega)/w_r^{in}(\omega)|\Delta_x$ and a Bogoliubov excitation of the form \eqref{eq:perturbation}. The amplitude $\beta_{r'}$ of the latter is inferred by integrating over its momentum distribution
\begin{equation}
    \beta_{r'}(k) = U_k \psi_{r'}(k) - V_k^* \psi_{r'}(-k)^*,
\end{equation}
where $\psi_{r'}(k) = \mathcal{F}[\psi_{r'}(x)]$ is the Fourier transform of the respective out-going wave packet in the atomic basis. The elements of the scattering matrix at the excited frequency $\omega$ are then obtained as 
\begin{equation}
    \abs{\mathcal{M}_{r'r}(\omega)}^2 = \abs{\frac{\beta_{r'}}{\alpha_r}}^2,
\end{equation}
where $\alpha_r$ is the similarly inferred amplitude of the in-going mode.

\section{Time-dependent Bogoliubov simulations}
\label{appendix:AppBogotimedep}
The linear dynamics of the small spin fluctuation field on top of the condensed (mean-field) component in the system is described by the corresponding time-dependent Bogolilubov theory~\cite{Castin1998TimeTraps,Butera2021PartCr}. This formalism generalizes the standard Bogoliubov theory introduced in Appendix~\ref{sec:App_A} to the case of a time-dependent background mean-field component. In the symmetric configuration considered here, both components are described by the single order parameter $ \psi_0(x,t) = \sqrt{2}\psi_\uparrow(x,t) = \sqrt{2}\psi_\downarrow (x,t)$, normalized to the total number of particles $N$. This evolves in time according to the Gross-Pitaevskii equation:
\begin{equation}
i\hbar\frac{\partial\psi_0(x,t)}{\partial t} = \hat{H}_{d} \psi_0(x,t)
\label{Eq:GPE_App}
\end{equation}
where
\begin{equation}
    \hat{H}_{d} \equiv -\frac{\hbar^2\partial_x^2}{2m} +\frac{(g+g_{\uparrow\downarrow})}{2} |\psi_0(x,t)|^2+V(x,t)-\frac{\hbar\Omega(x,t)}{2}
\end{equation}
is the Gross-Pitaevskii Hamiltonian. The evolution of the spin modes is governed instead by the Bogoliubov-de Gennes equations that, in the general time-dependent configuration, read as:
\begin{equation}
	i\hbar\frac{d}{dt}
	\begin{pmatrix}
    U\\
    V
  \end{pmatrix}=\mathcal{L}
  	\begin{pmatrix}
    U\\
    V
  \end{pmatrix} 
  =
  \begin{pmatrix}
    \mathcal{L}_{UU}      & \mathcal{L}_{UV}\\
    \mathcal{L}_{VU}    &  \mathcal{L}_{VV}
  \end{pmatrix}
    \begin{pmatrix}
    U\\
    V
  \end{pmatrix}.
  \label{Eq:BogOperator}
\end{equation}
Here, we defined the (operator-valued) components of the Bogoliubov operator $\mathcal{L}$ as:
\begin{align}
	\mathcal{L}_{UU}& = \hat{H}_d + \frac{\kappa(x,t)}{2} |\psi_0(x,t)|^2 + \hbar\Omega(x,t) - \mu,\nonumber\\
	\mathcal{L}_{UV}& = \frac{\kappa(x,t)}{2}  \psi_0^2(x,t),\nonumber\\
	\mathcal{L}_{VU}& = -\mathcal L_{UV}^* %= -[{\kappa}(x,t)/{2}] \big(\psi_0^*(x,t)\big)^2,\nonumber
 \\
        \mathcal{L}_{VV}& = - \mathcal{L}_{UU}.\label{Eq:BogoComponents}
\end{align}
Being quadratic in the field fluctuations, the Bogoliubov theory models the free evolution of the elementary excitation on top of the condensate, and is therefore able to describe both spontaneous and stimulated parametric emission processes, from which the Hawking radiation originates.
%as well as the cosmological particle creation~\cite{} and the dynamical Casimir effect~\cite{},

The numerical setup used for the time-dependent simulations presented in Sec.\ref{sec:GPtimedep} of the main text consists of an uniform, one-dimensional atomic condensate of length $L$, in a ring-shaped configuration (that is, we impose periodic boundary conditions). Specifically, we performed the simulations by using a system of length $L = 1895\,\xi_u$, discretized by using a numerical grid comprising $N_x = 2600$ points ($\delta x = L/N_x \approx 0.75 \xi_u$). The condensate initially flows subsonically with a certain momentum $q$, as described by the order parameter introduced in Eq.~\eqref{eq:hawking_stationarystate} in the main text. The Bogoliubov operator $\mathcal{L}$ defined in Eqs.~\eqref{Eq:BogOperator} and \eqref{Eq:BogoComponents} is diagonalised in such an initial configuration, so that the spin eigenmodes and the corresponding spectrum are obtained. As described in Sec.~\ref{sec:spinsonic}, the black hole for the spin fluctuations is then dynamically created by modulating both in time and in space the spin interaction strength $\kappa$ and the strength of the Rabi coupling $\Omega$ between the two components. Notice that, due to the periodic boundary conditions, also a white hole appears in the system, which we locate on the opposite side of the ring ($x=\pm L/2$) respect to the black hole ($x=0$) to avoid interference effects between the two horizons. 

In order to keep the mean-field component stationary, we modify the external potential in such a way that the chemical potential of the system remains constant, that is we prescribe: $V(x,t) - \hbar\Omega(x,t)/2 = \text{const.}$ The values of $\Omega(x,t)$, $\kappa(x,t)$ and $V(x,t)$ are %set both in the upstream (u) and downstream (d) regions of the ring, independently. These are 
modulated in time and space according to the following profiles:
\begin{align}
    \kappa_l(t) &= \frac{\kappa_{l,i}+\kappa_{l,f}}{2} + \left(\frac{\kappa_{l,f}-\kappa_{l,i}}{2}\right) \tanh \left(\frac{t-t_0}{\sigma_t}\right)
    %= \kappa_{l,i} - \left(\frac{\kappa_{l,i}-\kappa_{l,f}}{2}\right) \left[\tanh \left(\frac{t-t_0}{\sigma_t}\right)+1\right]
    \qquad(l=u,d),\label{Eq:TimeProfile}\\
    \kappa(x,t) &= \frac{\kappa_u(t) + \kappa_d(t)}{2} + \left(\frac{\kappa_u(t) - \kappa_d(t)}{2}\right)\left[ \tanh\left(\frac{x+L/2}{\sigma_x}\right) + \tanh\left(\frac{x-L/2}{\sigma_x}\right) - \tanh\left(\frac{x}{\sigma_x}\right) \right] 
    %\kappa_d(t) + \left(\frac{\kappa_u(t) - \kappa_d(t)}{2}\right)\left[1 + \tanh\left(\frac{x+L/2}{\sigma_x}\right) + \tanh\left(\frac{x-L/2}{\sigma_x}\right) - \tanh\left(\frac{x}{\sigma_x}\right) \right].
    \label{Eq:SpatialProfile}
\end{align}
Similar profiles have also been used for the Rabi frequency and the external potential. In Eqs.~\eqref{Eq:TimeProfile} and \eqref{Eq:SpatialProfile}, $\sigma_{t/x}$ give information on the time/spatial scale on which the variation happens, while $t_0$ is the time instant at which the time modulation is centered. In physical terms, $\sigma_x$ represents the horizon's width and $t_0 = 90 m c_u^2/\hbar$ is the instant at which the horizon is generated.

The mean-field component of the system and each Bogoliubov mode are propagated in time by numerically solving Eqs.~\eqref{Eq:GPE} and \eqref{Eq:BogOperator}, respectively\footnote{Notice that, since the chemical potential is kept constant, the mean-field component remains stationary and equal to its initial configuration through the whole evolution.}. To this end, we adopt a symmetric fourth order finite difference scheme~\cite{fornberg1988fdm} for evaluating spatial derivatives, and the fourth order Runge-Kutta Method for propagating the solution in time. Correlations functions at the time $t$ are then calculated as described in Eqs.~\eqref{eq:G33definition}, \eqref{eq:G22definition} and \eqref{eq:G23definition}, by using the spin fluctuation field decomposition in Eq.~\eqref{eq:fieldpert}, with the eigenmodes $U(x,t),V(x,t)$ evaluated at the time $t$.

%\begin{figure*}[t!]
%    \centering
%    \includegraphics[width=\linewidth]{Figures/DD_Gapless_Galess_dx4.png}\\
%    \includegraphics[width=\linewidth]{Figures/PD_Gapless_Galess_dx4.png}\\
%    \caption{horizon length $\Delta x/\xi_s = 4$}
%    \label{fig:gapless_time_dep_dx4}
%\end{figure*}

\section{The effects of a smooth horizon: Hawking temperature and dynamical instabilities}
\label{appendix:horizoneffect}

As already pointed out, time-dependent simulations not only validate the semi-analytical results obtained for the idealized step configuration, but also allow to investigate the effect of a horizon of finite width on both spontaneous and stimulated Hawking radiation. 

As seen in Fig.~\ref{fig:S_omega_ideal}(a), the smoothened horizon features a lower reflectivity $|\mathcal M_{uu}(\omega)|^2$ of in-going $u$-modes than the infinitely sharp case, most notably in the limit $\omega \to \omega_*$. More generally, the limit $\omega \to \infty$ corresponds to a regime in which the chemical potential is locally constant with respect to the length scale set by momentum of the in-going wave packet, resulting in a vanishing reflectivity.
Regarding the Hawking process in particular, we find,
in analogy with results from previous literature on single-component BECs \cite{macher2009black}, that the main effect of a finite horizon width $\sigma_x$ is
a weakened signature of the Hawking process with respect to the semi-analytical prediction. This can be seen both in Fig.~\ref{fig:S_omega_ideal} and by comparing Fig.~\ref{fig:gapped} with Fig.~\ref{fig:gapped_time_dep_dx2_dx4}(c). 

\begin{figure}[bt]
    \includegraphics[width = \linewidth]{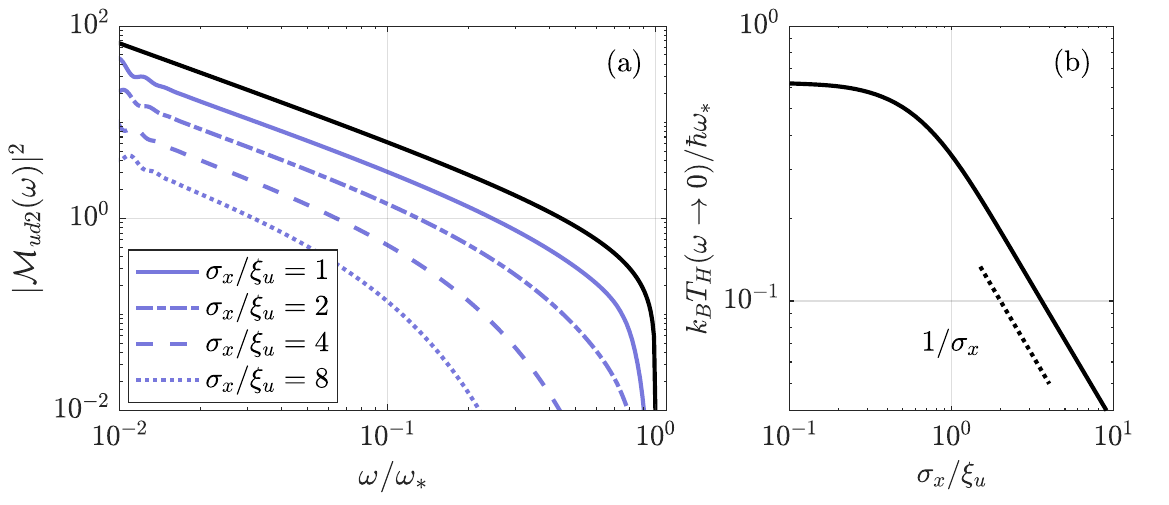}
    \caption{Extended data for Fig.~\ref{fig:S_omega_ideal}. (a) Frequency dependence of $|\mathcal{M}_{ud2}|^2$ in an uncoupled mixture ($\kappa_d/\kappa_u = 0.25$, $\Omega_d = 0$), inferred through GP scattering simulations for different smoothness profiles \eqref{eq:smooth_horizon_profile} of the spin-sonic horizon. The black line indicates the semi-analytical Bogoliubov prediction. (b) Analog Hawking temperature as defined in \eqref{eq:hawkingspectrum_gapless} as a function of the horizon smoothness.}
    \label{fig:horizon_width}
\end{figure}

Figure~\ref{fig:horizon_width}(a) illustrates for the gapless case ($\Omega_d=0$) how the Hawking signal decays with increasing smoothness $\sigma_x$ of the spin-sonic horizon.
Through the emission spectrum \eqref{eq:hawkingspectrum_gapless}, this can be summarized in terms of the Hawking temperature $T_H(\omega)$ which, in the absence of a coherent coupling, is frequency independent in the limit $\omega \to 0$. 
As shown in Fig.~\ref{fig:horizon_width}(b), the analog Hawking temperature decays as $T_H \sim \sigma_x^{-1}$. This result is consistent with the prediction \cite{visser1998hawking, macher2009black}
\begin{equation}
    k_B T_H = \frac{\hbar }{2\pi } \frac{\partial(c-v)}{\partial x} \bigg|_{x=0}
    \label{eq:analogTH}
\end{equation}
of the hydrodynamic approximation using the horizon profile \eqref{eq:smooth_horizon_profile} for which $\partial c / \partial x = (c_u^2 - c_d^2)/4\sigma_x$. For an increasingly steep horizon, the Hawking temperature saturates to a finite value $k_BT_H(\omega \to 0)/\hbar\omega_* \simeq 0.68$, in agreement with the semi-analytical prediction for a sharp horizon in Sec. \ref{sec:THgreybody}.
If $\sigma_x \sim 2\xi_u$ (the value we used in the main text), $T_H$ is predicted to be roughly half of such maximum value, as illustrated in Fig.~\ref{fig:Hspectrum}(b). Consistently, the phase-phase correlation signal in Fig.~\ref{fig:gapped_time_dep_dx2_dx4} is approximately a factor 2 smaller than the semi-analytical prediction in Fig.~\ref{fig:gapped}. 

%--------------------
%--------------------

\begin{figure*}[ht]
    \centering
    \includegraphics[width=\linewidth]{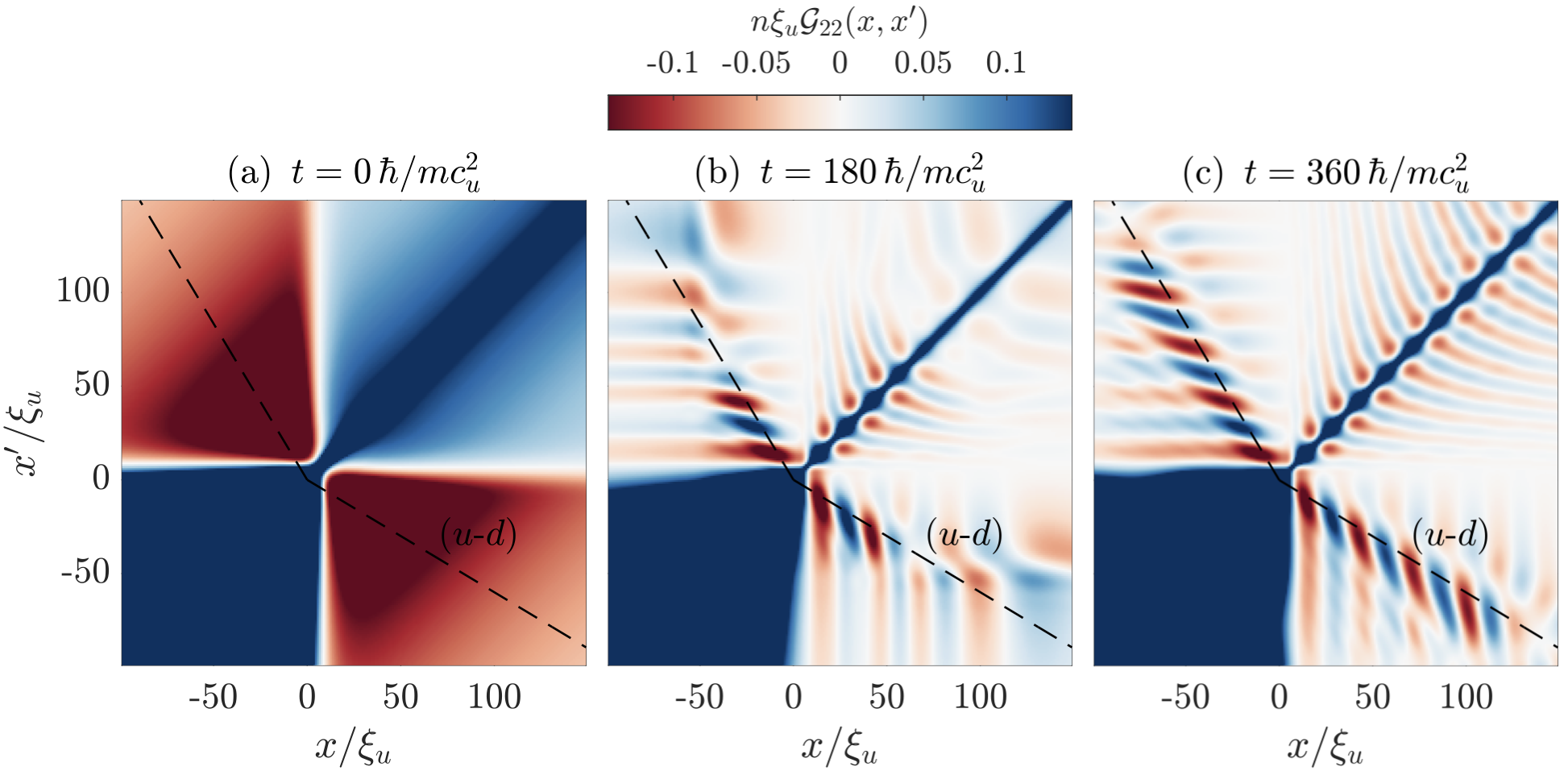}
    \caption{Time evolution of equal-time phase-phase correlation functions, in the presence of a coherent coupling in the downstream region. In the two cases, snapshots of the correlations have been taken at the same time instants described in Fig.~\ref{fig:gapless_time_dep_dx2}.  These results provide the time dependent scenario for the stationary solution given in Fig.~\ref{fig:gapped}, but with a very smooth horizon of width $\sigma_x = 3.8\xi_u$. }
    \label{fig:gapped_smoothhorizon4}
\end{figure*}

The size of the horizon not only affects the strength of the Hawking signal, because of its connection to the temperature of the emission, but it can also modify the spatial structure of the correlations if it is much larger than the healing length. An example is shown in Fig.~\ref{fig:gapped_smoothhorizon4} which reports a phase-phase correlation signal analogous to Fig.~\ref{fig:gapped_time_dep_dx2_dx4}(c), obtained with an even larger $\sigma_x \sim 3.8\,\xi_u$: in addition to a reduction of the intensity of the signal by a factor 10, we also observe that the checkerboard pattern in the $x,x'>0$ sector becomes a "ribcage" structure.  

As can be seen in Fig.~\ref{fig:horizon_width}, if the coherent coupling is vanishing, the regime $\sigma_x \lesssim \xi_u$ can be reached without numerical issues and the semi-analytical result for the step configuration is recovered in the limit $\sigma_x \ll \xi_u$ (a similar result was obtained for a non-vanishing and uniform Rabi frequency $\Omega_d=\Omega_u$ on both sides of the horizon~\cite{fernandes2024thesis}). Remarkably, this is not the case if $\Omega$ has different asymptotic values on the two sides of the horizon, a configuration that we considered in Figs. \ref{fig:gapped} and \ref{fig:gapped_time_dep_dx2_dx4}: too a sharp variation of the Rabi frequency in space ($\sigma_x \lesssim 1.5\xi_u$), even if properly compensated by an external potential, results in the presence of dynamically unstable spin modes in the GP numerical simulation, whose dynamics prevents the observation of the Hawking signal. In particular, the checkerboard pattern in the downstream region of Fig.~\ref{fig:gapped}(c) grows indefinitely in time as a power-law, whose exponent is related in a non-trivial way to the magnitude of the Rabi frequency $\Omega$. On the other hand, the GP simulations confirm the dynamical stability of spin-sonic horizon for sufficiently smooth horizon with $\sigma_x \gtrsim 1.5\xi_u$. 
A further analysis of physical origin of these instabilities will be the subject of future works.

%By doubling the value of $\sigma_x$, we have verified that the structures of $G_{33}(x,x')$ and $G_{23}(x,x')$ are weakly sensitive to the horizon's width, in the case of the uncoupled system (results not shown). The only noticeable difference is a weaker Hawking signal, as expected due to the corresponding lower Hawking temperature (cfr. Sec~\ref{}). In the coupled configuration instead, we observe that the structure of the correlations is affected by the size of the black hole; such a behavior is evident, for example, in Fig.~\ref{fig:gapped_time_dep_dx2_dx4} by comparing $G_{22}(x,x')$ obtained with a horizon of size $\sigma_x \approx 1.9\,\xi_s$ [panels (a-c)], and $\sigma_x \approx 3.8\,\xi_s$ [panels (d-f)].
%Related to this, it is worth noticing that both the time-dependent Bogoliubov and GP simulations reveal an instability of the spin fluctuations in the gapped configuration, when the size of the horizon is $\sigma_x \lessapprox 1.5 \xi_s$. As result of such an instability, the checkerboard signal in the downstream region grows indefinitely in time as a power-law, whose power is related in a non-trivial way to the mangitude of the coherent coupling $\Omega$. This effect is detrimental for the physics investigate, since it overwhelms the Hawking signal. Such an intriguing behaviour will be subject of further investigation in a later work.